# Numerical Modeling of $Cu_2MnSnS_4$/$FeSi_2$ Dual-Absorber Solar Cell Achieving High Efficiency

Hasib Md Abid Bin Farid [a,1], and Md Tashfiq Bin Kashem [a,1,*]

[a] Department of Electrical and Electronic Engineering, Ahsanullah University of Science and Technology, Dhaka – 1208, Bangladesh

[1] Equal contribution

[*] Corresponding author: E-mail address: tashfiq.eee@aust.edu

**Abstract:** Dual-absorber solar cells represent a promising approach to surpass the efficiency limit of single-junction devices by extending spectral absorption and minimizing thermalization losses. Among earth-abundant thin-film materials, kesterites have attracted considerable interest; however, the well-studied $Cu_2ZnSnS_4$ (CZTS) continues to face challenges related to antisite disorder and secondary phase formation. Replacing Zn with Mn in $Cu_2MnSnS_4$ (CMTS) mitigates these limitations, improving cation ordering and electronic quality while maintaining favorable optical properties. Yet, despite its potential, CMTS remains largely unexplored in multi-absorber configurations—only one prior study has reported a CMTS-based dual-absorber device. In this work, we present a comprehensive numerical investigation of a CMTS/$FeSi_2$ dual-absorber thin-film solar cell using the one-dimensional solar cell capacitance simulator (SCAPS-1D). $FeSi_2$, with its narrow band gap (~0.87 eV) and strong near-infrared absorption, serves as an ideal bottom absorber to complement CMTS, enabling broader spectral utilization. The study systematically examines the effects of geometrical, electronic, and interfacial parameters on carrier transport, and overall device performance. The optimized device delivers an impressive power conversion efficiency of 34.9%, with $V_{OC}$ = 0.79 V, $J_{SC}$ = 51.07 mA/cm², and a fill factor of 85.91%. These findings demonstrate that integrating $FeSi_2$ with CMTS not only enhances carrier collection and spectral harvesting but also establishes a new pathway toward high-efficiency, sustainable, and environmentally benign thin-film photovoltaics.

**Keywords:** Thin-film solar cell; Dual-absorber solar cell; CMTS; $FeSi_2$; SCAPS-1D

## 1. Introduction

The growing global demand for energy, coupled with the adverse environmental impact of fossil fuel use, has accelerated the transition toward renewable energy technologies. Among these, solar photovoltaics (PV) have emerged as one of the most viable solutions because sunlight is abundant, clean, and widely available [1], [2]. Currently, crystalline silicon (c-Si) dominates the PV market with record efficiencies exceeding 26% [3]–[5]. However, the high energy cost of wafer processing, the requirement for relatively thick absorbers, and the fundamental Shockley–Queisser (SQ) efficiency limit constrain further progress in c-Si technology [4], [6]–[9].

Thin-film solar cells offer a promising pathway to overcome these challenges. They consume less material, can be produced at lower cost, and enable flexible device designs [4], [7], [10], [11]. Established technologies such as cadmium telluride (CdTe) and copper indium gallium selenide (CIGS) have demonstrated power conversion efficiencies above 22% [12]–[15]. However, both face sustainability challenges: CdTe contains toxic cadmium, while CIGS depends on scarce elements such as indium and gallium [7], [10], [16]–[19]. These issues raise concerns about large-scale deployment and highlight the need for thin-film absorbers that are efficient, environmentally benign, and earth-abundant [10], [20].

Kesterite compounds have gained attention as promising alternatives. $Cu_2ZnSnS_4$ (CZTS), in particular, is attractive due to its direct band gap near 1.5 eV and high absorption coefficient (> $10^4$ cm$^{-1}$) [10], [16], [21]–[26]. Despite these favorable properties, persistent problems such as structural defects - particularly Cu-Zn antisite disorder due to the similar ionic radii of $Cu^+$ and $Zn^{2+}$, and the formation of secondary phases such as ZnS have limited device efficiencies to around 13%, well below the theoretical potential [6], [10], [16], [20], [27]–[31].

Cation substitution has emerged as a promising route to address these challenges [10], [16], [20], [28], [32], [33]. In particular, substituting zinc with manganese yields $Cu_2MnSnS_4$ (CMTS), which preserves strong optical absorption while enabling bandgap tuning in the 1.3–1.6 eV range [7], [20], [34]–[39]. The incorporation of $Mn^{2+}$ in place of $Zn^{2+}$ effectively reduces intrinsic cation disorder and suppresses antisite defects such as Cu–Zn exchange, owing to the larger ionic radius of Mn [4], [8], [16]. This enhanced cation ordering minimizes the formation of deep recombination centers and improves electronic quality. Furthermore, the substitution of Zn with Mn promotes phase purity by suppressing the formation of stable secondary phases such as ZnS, which are detrimental to carrier transport. Owing to the comparatively lower stability of MnS, this substitution facilitates the formation of single-phase CMTS [20], [40]. Beyond these structural and electronic advantages, manganese is inexpensive, non-toxic, and earth-abundant [8], [34], [40]. Combined with its favorable bandgap, low reflectivity, high absorption coefficient ($>10^4$ cm$^{-1}$ in the visible range), and structural similarity to well-studied kesterites, CMTS stands out as a promising absorber for high-performance thin-film solar cells [7], [37], [41].

CMTS has been synthesized using a wide variety of solution-based, vacuum-based, and electrochemical deposition techniques [34], [42]–[47]. However, despite these diverse fabrication processes, experimentally realized CMTS solar cells have so far delivered relatively low efficiencies—typically below 2% [16], [34], [37], [47]–[50]. This performance gap highlights the necessity of numerical simulation prior to further fabrication and optimization, as it enables the prediction of device behavior, identification of bottlenecks, and evaluation of design strategies. Software such as SCAPS-1D (Solar Cell Capacitance Simulator) has become an essential tool in this regard [51]–[55]. So far, simulations of CMTS-based solar cells have predicted efficiencies exceeding 20% under optimized conditions [4], [7], [10], [16], [56], with some reports projecting values above 30% [7]. These results strongly suggest that the poor experimental efficiencies reported so far are not due to intrinsic material limitations but rather to fabrication challenges. Simulations also allow the exploration of novel device architectures that are difficult to achieve experimentally in the early stages of material development. This predictive capability makes simulation not just a supportive tool but an essential step in advancing new absorber materials toward practical implementation.

However, even with the encouraging efficiencies projected for CMTS solar cells through numerical simulations, like all single-junction devices, they remain intrinsically bound by the Shockley–Queisser (SQ) limit, which caps the maximum theoretical efficiency at ~33% [6], [57], [58]. This fundamental restriction arises from two primary loss mechanisms: the inability to absorb photons with energies below the band gap and the thermalization of carriers generated by higher-energy photons. Together, these losses can account for nearly 55% of the incident solar energy, representing a major bottleneck in the performance of single-bandgap photovoltaic systems [59]–[61]. While several advanced concepts—such as multijunctions [61], multiple spectrums [62], and multiabsorption techniques [60]—have been proposed to overcome these losses, most remain in the conceptual or laboratory stage due to complex fabrication requirements [57].

A more practical and technologically feasible strategy is the adoption of dual-absorber architectures, where two absorber layers with complementary band gaps are stacked to broaden spectral coverage [57]. In such systems, the wide-bandgap absorber efficiently harvests high-energy photons while the narrow-bandgap layer captures lower-energy photons, thereby reducing thermalization and transmission losses simultaneously. This spectral complementarity not only improves carrier generation but also enhances photoelectric conversion efficiency by facilitating better utilization of the solar spectrum [57], [63], [64]. Moreover, band alignment between the two absorbers can create favorable energy gradients that promote directional carrier transport and suppress recombination at the interface. As a result, dual-absorber solar cells have been successfully demonstrated across several other thin film material systems, including CIGS/CZTSSe [65], [66], CZTGSe/Sb$_2$Se$_3$ [67], CZTS/Sb$_2$Se$_3$ [68][69], CZTSSe/CsPbI$_3$ [70], CZTS/Si [71][72], CIGS/CsSnI$_3$ [73], CIGS/MASnI$_3$ [74][75], CIGS/KSnCl$_3$ [75], and CIGS/CZTSe [76], showing notable efficiency gains over their single-junction counterparts.

To date, however, CMTS has rarely been explored in this context. To the best of the author's knowledge, only one study has reported a CMTS-based dual-absorber solar cell, employing $Sb_2Se_3$ as the secondary absorber, which achieved a power conversion efficiency of 12.86% [2]. This promising yet isolated result highlights both the potential and the underexplored nature of CMTS in multi-absorber configurations. Therefore, further investigation into its role within dual-absorber architectures—especially with other earth-abundant partners—could unlock new opportunities for high-efficiency, sustainable thin-film photovoltaics.

Among the potential candidates, $FeSi_2$ stands out as a material with remarkable optoelectronic and stability characteristics, though its synergy with CMTS has yet to be investigated. With a direct band gap of ~0.80–0.87 eV and an optical absorption coefficient exceeding $1\times10^5$ cm$^{-1}$, $FeSi_2$ is well-suited to capture the near-infrared (NIR) portion of the solar spectrum that wide-bandgap absorbers typically miss [77]–[82]. It exhibits excellent chemical stability, strong resistance to oxidation, humidity, and cosmic radiation, and can withstand high operating temperatures without degradation [77], [78], [83]–[85]. Importantly, $FeSi_2$ is composed of earth-abundant, non-toxic elements (Fe and Si), making it both environmentally friendly and low-cost [86]–[88]. Additional advantages include a long carrier diffusion length (~38 μm), mechanical compatibility with diverse substrates, and suitability for integration into silicon-based technologies [78], [85], [89]. Collectively, these attributes position $FeSi_2$ as a durable, sustainable, and efficient absorber material, particularly valuable as a bottom absorber in dual-absorber solar devices [81], [82], [90], [91].

In this study, SCAPS-1D simulations are employed to conduct a comprehensive evaluation of CMTS/$FeSi_2$ dual-absorber thin-film solar cells, explored here for the first time. The investigation considers a wide range of factors, including geometrical parameters (layer thicknesses), electronic properties (doping and defect densities), and interfacial or operational conditions (contact work functions, series and shunt resistances, temperature, and illumination intensity). By systematically varying these parameters, the study examines their influence on carrier transport, recombination, and overall device efficiency to identify optimal design conditions. The findings indicate that introducing $FeSi_2$ as a secondary absorber in the CMTS-based device leads to a notable improvement in photovoltaic performance, highlighting its promise for highly efficient and sustainable thin-film solar technologies.

## 2. Simulation methodology

All the simulations in this study were performed using SCAPS-1D (version 3.3.12), a numerical tool developed at the University of Ghent for modeling solar cells [92]–[94]. This software has demonstrated excellent consistency with experimentally reported results in previous studies [82], [95]–[101]. The program numerically solves the coupled one-dimensional Poisson and carrier continuity equations under steady-state conditions.

Poisson's equation:
$$\frac{\partial^2 \Phi(x)}{\partial x^2} = -\frac{q}{\epsilon}\left[p(x) - n(x) + N_D^+ - N_A^- \pm N_{def}(x)\right] \quad (1)$$

Continuity equation:
$$\begin{cases} \text{Electron: } \frac{\partial n}{\partial t} = \frac{1}{q}\frac{\partial J_n}{\partial x} + (G_n - R_n) & (2) \\ \text{Hole: } \frac{\partial p}{\partial t} = -\frac{1}{q}\frac{\partial J_p}{\partial x} + (G_p - R_p) & (3) \end{cases}$$

Here, $\Phi$ is the electrostatic potential, $q$ is the electronic charge, $\epsilon$ is the permittivity of the material, $p$ and $n$ are hole and electron density respectively, $N_D^+$ and $N_A^-$ are donor and acceptor density respectively, $N_{def}$ is the defect density, $G$ and $R$ are generation and recombination rate respectively, and $J$ is the current density. Carrier transport is governed by drift-diffusion mechanisms, where the current densities $J_n$ and $J_p$ for electrons and holes are described as:

$$\text{Charge transport equation:} \begin{cases} \text{Electron:} J_n = qn\mu_n E + qD_n \frac{\partial n}{\partial x} & (4) \\ \text{Hole:} J_p = qp\mu_p E - qD_p \frac{\partial p}{\partial x} & (5) \end{cases}$$

where $\mu$ is the mobility, $E$ is the electric field and $D$ is the diffusion coefficient. In all the equations, the subscripts, $n$ and $p$ correspond to electron and hole respectively. The optical absorption in each layer is calculated in SCAPS-1D using the 'E$_g$-sqrt' model [102]:

$$\alpha(h\upsilon) = \begin{cases} \left[\alpha_0 + \beta_0 \frac{E_g}{h\upsilon}\right] \sqrt{\frac{h\upsilon}{E_g} - 1}, & h\upsilon > E_g \\ 0, & h\upsilon < E_g \end{cases} \quad (6)$$

Here, $\alpha$ is the absorption coefficient of the material, $\alpha_0$ and $\beta_0$ are model constants, $E_g$ is the energy bandgap of the material, $h\upsilon$ is photon energy where $h$ is Planck's constant, and $\upsilon$ is photon frequency. Device performance metrics such as open-circuit voltage (V$_{OC}$), short-circuit current density (J$_{SC}$), fill factor (FF), and power conversion efficiency (PCE) are extracted from the J-V characteristics [103]:

$$V_{OC} = \frac{kT}{q} \ln\left(\frac{J_{SC}}{J_0} + 1\right) \quad (7)$$

$$FF = \frac{V_m J_m}{V_{OC} J_{SC}} \quad (8)$$

$$PCE = \frac{V_{OC} J_{SC} FF}{P_{in}} \quad (9)$$

Where $k$ is the Boltzmann constant, $J_0$ is the reverse saturation current density, $V_m$ and $J_m$ refer to the voltage and current density respectively at the maximum power point, and $P_{in}$ is the input power.

## 3. Device structure and material parameters

The simulated device architecture in this work consists of the configuration: front contact / ITO / CdS / CMTS / FeSi$_2$ / Cu$_2$O / back contact, as schematically illustrated in Fig. 1(a). Here, indium tin oxide (ITO) serves as the transparent conducting oxide (TCO) and window layer, ensuring high optical transmittance and efficient carrier collection at the front interface [104][105]. The CdS layer is employed as the electron transport layer (ETL) because of its wide band gap (~2.4 eV), suitable conduction band alignment with solar cell absorbers, and proven ability to minimize interface recombination losses [106][107]. CdS has been widely adopted in CZTS, CIGS, and CMTS thin-film solar cells owing to its excellent lattice matching and chemical stability, which facilitate efficient charge extraction and transport toward the front electrode [7], [108]–[110].

The CMTS layer functions as the primary absorber, responsible for generating electron–hole pairs under illumination, while the FeSi$_2$ layer acts as the secondary absorber to extend the spectral response into the near-infrared (NIR) region, enhancing overall light-harvesting efficiency. The Cu$_2$O layer is introduced as the hole transport layer (HTL) due to its high hole mobility, suitable valence band position, and good chemical compatibility with chalcogenide absorbers [111][112]. In addition, Cu$_2$O has been demonstrated to form stable, low-resistance contacts with several thin-film absorber materials, thereby facilitating efficient hole extraction and minimizing recombination at the back interface [10], [110], [113], [114].

Aluminum (Al) and gold (Au), with work functions of approximately 4.3 eV and 5.1 eV, respectively [115]–[117], are chosen as the front and back contact metals to ensure favorable band alignment with ITO and Cu$_2$O, facilitating efficient electron and hole collection. All material and interface parameters such as thickness, bandgap, electron affinity, dielectric constant, mobility, and defect densities, were specified based on literature values or estimated where unavailable, as summarized in Tables 1 and 2. Recombination mechanisms included in the simulation are radiative,

Table 1. Input parameters of the materials: band gap ($E_g$), electron affinity ($\chi$), relative dielectric permittivity ($\varepsilon_r$), effective density of states in the conduction band ($N_c$), effective density of states in the valence band ($N_v$), electron mobility ($\mu_n$), hole mobility ($\mu_h$), shallow uniform acceptor density ($N_A$), shallow uniform donor density ($N_D$), and defect density ($N_t$)

| Parameters | ITO | CdS | CMTS | FeSi$_2$ | Cu$_2$O |
|---|---|---|---|---|---|
| Thickness (nm) | 300 | 50 | 100 | 300 | 100 |
| $E_g$ (eV) | 3.5 | 2.4 | 1.60 | 0.87 | 2.2 |
| $\chi$ (eV) | 4 | 4.2 | 4.35 | 4.16 | 3.2 |
| $\varepsilon_r$ | 9 | 10 | 9 | 22.5 | 7.1 |
| $N_c$ (1/cm$^3$) | $2.2 \times 10^{18}$ | $2.2 \times 10^{18}$ | $2.2 \times 10^{18}$ | $5.6 \times 10^{19}$ | $2.5 \times 10^{18}$ |
| $N_v$ (1/cm$^3$) | $1.8 \times 10^{19}$ | $1.8 \times 10^{19}$ | $1.8 \times 10^{19}$ | $2.08 \times 10^{19}$ | $1.5 \times 10^{19}$ |
| $\mu_n$ (cm$^2$/Vs) | 20 | 100 | 0.16 | 100 | 100 |
| $\mu_h$ (cm$^2$/Vs) | 10 | 25 | 0.16 | 20 | 80 |
| $N_D$ (1/cm$^3$) | $10^{21}$ | $10^{18}$ | 0 | 0 | 0 |
| $N_A$ (1/cm$^3$) | 0 | 0 | $10^{16}$ | $10^{19}$ | $10^{16}$ |
| Defect type | Neutral | Acceptor | Neutral | Donor | Neutral |
| $N_t$ (1/cm$^3$) | $10^{14}$ | $10^{14}$ | $10^{11}$ | $10^{14}$ | $10^{15}$ |
| Reference | [99],[118] | [119]–[121] | [2] | [77],[85],[89] | [10] |

ITO = Indium-doped tin oxide; CdS = Cadmium sulfide; CMTS = Copper manganese tin sulfide; FeSi$_2$ = iron disilicide; Cu$_2$O = cuprous oxide

Table 2. Input parameters for interface defects

| Parameters | ETL/CMTS interface | CMTS/FeSi$_2$ interface | FeSi$_2$/HTL interface |
|---|---|---|---|
| Defect type | Neutral | Neutral | Neutral |
| Electron capture cross section (cm$^2$) | $10^{-19}$ | $10^{-19}$ | $10^{-15}$ |
| Hole capture cross section (cm$^2$) | $10^{-19}$ | $10^{-19}$ | $10^{-15}$ |
| Energetic distribution | Single | Single | Single |
| Reference for defect energy level | Above the highest $E_V$ | Above the highest $E_V$ | Above the highest $E_V$ |
| Energy with respect to reference (eV) | 0.6 | 0.6 | 0.6 |
| Total density (cm$^{-2}$) | $10^{10}$ | $10^{10}$ | $10^{10}$ |

Auger, and Shockley-Read-Hall (SRH) recombination. Unless otherwise specified, all SCAPS-1D simulations are performed under standard AM1.5G illumination with an incident intensity of 1000 W/m² and an operating temperature of 300 K. The energy band alignment of the complete device is shown in Fig. 1(b).

## 4. Results and discussion

### 4.1. Single versus dual-absorber solar cell

To investigate the influence of incorporating $FeSi_2$ as the second absorber layer on the photovoltaic performance, simulations were conducted for both a single absorber cell using only CMTS and a dual absorber cell incorporating CMTS and $FeSi_2$, under identical conditions. Key output characteristics such as current–voltage (J–V) behavior, quantum efficiency (QE), and performance metrics were analyzed (Fig. 2 and Table 3).

Table 3 shows that, in the device with only CMTS as the absorber (bandgap ~ 1.6 eV), $V_{OC}$ is high (~1.13 V), while $J_{SC}$ is 19.31 mA/cm². Upon adding $FeSi_2$ (bandgap ~0.87 eV) as the secondary absorber, $V_{OC}$ decreases to 0.7 V, whereas $J_{SC}$ rises substantially to 51.04 mA/cm². FF experiences a modest reduction from 81.25% to 75.1%, and the overall PCE increases from 17.67% to 26.97%. The QE spectra further reveal that the CMTS-only device exhibits a sharp drop in QE beyond ~780 nm, whereas the dual absorber structure maintains high QE well beyond 1100 nm, indicating effective absorption of longer-wavelength photons by $FeSi_2$.

These results arise from the complementary roles of the two absorbers. The reduced $V_{OC}$ in the dual absorber cell is expected, as the overall quasi-Fermi level splitting is limited by the narrower-bandgap $FeSi_2$ [73], [123]. The large increase in $J_{SC}$ is directly supported by the extended QE response [124] (Fig. 2b), photons not absorbed by CMTS are efficiently captured by $FeSi_2$, generating additional photocurrent (Fig. 3) while maintaining effective carrier extraction across the heterojunction [73], [81], [91], [123], [125], [126]. The slight reduction in FF reflects slight resistive and recombination penalties introduced by the additional layer, yet the device retains good diode behavior. Consequently, the obtained higher PCE demonstrates the advantage of dual absorbers for improved spectral coverage and photocurrent generation.

Building on the demonstrated superior performance of the CMTS/$FeSi_2$ dual-absorber configuration, subsequent simulations were performed to investigate how key material and device parameters influence carrier dynamics and overall device efficiency.

### 4.2. CMTS thickness and doping density

To assess how variations in the CMTS absorber properties affect overall device performance, the layer thickness and acceptor doping density were systematically varied while keeping all other parameters constant (Fig. 4a–d). Figures show that, $V_{OC}$ increases with higher acceptor doping, with a more pronounced improvement at larger thicknesses. $J_{SC}$ decreases with increasing doping density and shows a thickness-dependent behavior: at low doping ($10^{16}$–$10^{17}$ cm⁻³), thicker layers reduce $J_{SC}$, whereas at higher doping densities, thicker CMTS increases $J_{SC}$. FF improves with higher doping and exhibits minor thickness dependence: slight reductions occur in thicker CMTS at low doping, while it remains stable or slightly improves with thickness at higher doping. PCE reflects these combined effects: at fixed thickness, it decreases with increasing doping due to the drop in $J_{SC}$, while increasing thickness improves PCE at moderate-to-high doping densities.

Table 3. Comparison of performance parameters of single and dual absorber solar cells

| Device structure | $V_{OC}$ (V) | $J_{SC}$ (mA/cm²) | FF (%) | PCE (%) |
| --- | --- | --- | --- | --- |
| Al / ITO / CdS / CMTS / $Cu_2O$ / Au (Single absorber) | 1.126 | 19.31 | 81.25 | 17.67 |
| Al / ITO / CdS / CMTS / $FeSi_2$ / $Cu_2O$ / Au (Dual absorber) | 0.704 | 51.04 | 75.09 | 26.97 |

These trends can be attributed to the interplay of electric field strength, optical absorption, and bulk recombination. Higher CMTS doping enhances hole concentration and strengthens the built-in field, which promotes charge separation and reduces interface and depletion-region recombination, thereby increasing $V_{OC}$ and improving FF. However, excessive doping can narrow the depletion region, limiting carrier collection and reducing $J_{SC}$. Increasing absorber thickness extends the optical path, enhancing photon absorption and photocarrier generation, which increases $J_{SC}$, but excessive thickness can elevate bulk recombination, slightly lowering $V_{OC}$ and FF unless compensated by strong internal fields. Optimal device performance is achieved by balancing sufficient CMTS thickness with moderate-to-high acceptor doping, minimizing recombination while maximizing field-assisted carrier collection [16], [127]–[129].

### 4.3. FeSi$_2$ thickness and doping density

The effects of the doping density and thickness of the secondary absorber, FeSi$_2$ on the device performance are shown in Fig. 5. $V_{OC}$ exhibits a nonlinear dependence on FeSi$_2$ acceptor doping: for a given thickness, $V_{OC}$ increases modestly as the doping density rises up to ≈$10^{20}$ cm$^{-3}$ [81], with a more pronounced rise beyond this range; at fixed doping, increasing FeSi$_2$ thickness leads to a slight decline in $V_{OC}$. $J_{SC}$ displays minimal sensitivity to acceptor doping [81] but increases consistently with FeSi$_2$ thickness [81]. FF improves with increasing acceptor doping, and shows only modest thickness dependence—small reductions in thicker FeSi$_2$ at low doping and stability or slight improvement at higher doping [81]. Consequently, PCE rises with both higher FeSi$_2$ doping and thickness: doping boosts $V_{OC}$ and FF within an optimal window, while thickness increases photon harvesting and thus $J_{SC}$ [81].

As in the CMTS absorber analysis, these trends are governed by the competing effects of electric-field strength, optical absorption and bulk recombination. Increasing acceptor doping strengthens band bending and the built-in field, which promotes quasi-Fermi level splitting (raising $V_{OC}$) and improves conductivity (raising FF) but can also narrow the depletion width and reduce $J_{SC}$. However, within the studied doping range, this reduction in $J_{SC}$ was negligible [81]. Increasing absorber thickness enhances light absorption and photogeneration (increasing $J_{SC}$), yet excessive thickness raises bulk recombination when the layer exceeds the minority-carrier diffusion length and can slightly reduce $V_{OC}$ and FF unless compensated by stronger fields or improved conductivity through higher doping. Thus, an optimum exists where moderate-to-high FeSi$_2$ doping combined with sufficient thickness balances enhanced absorption and field-assisted collection against recombination and resistive losses, yielding the improvements in PCE [81]. The insights gained here, together with those from the previous section, form the basis for the subsequent analysis of coupled CMTS–FeSi$_2$ absorber interactions to understand how their electronic and optical properties jointly determine the overall device performance.

### 4.4. Combined effect of CMTS and FeSi$_2$ thickness

To evaluate the coupled influence of both absorber layers on device performance, the CMTS and FeSi$_2$ thicknesses were varied simultaneously while keeping other parameters constant (Fig. 6). From the simulations, it was observed that higher $V_{OC}$ was observed when the CMTS thickness exceeded 0.7 μm while keeping the FeSi$_2$ layer thin (< 0.2 μm). In contrast, $J_{SC}$ was maximized with a thinner CMTS layer (< 0.5 μm) and a thicker FeSi$_2$ layer (> 0.5 μm), as this configuration allowed better harvesting of longer-wavelength photons absorbed by FeSi$_2$. The highest FF occurred when the CMTS layer was very thin (< 0.1 μm) and FeSi$_2$ was thick (> 0.7 μm), likely due to enhanced charge extraction and reduced resistive losses. Consequently, the optimal PCE was achieved when CMTS thickness was below 0.2 μm and FeSi$_2$ exceeded 0.5 μm, representing an optimal trade-off among light absorption, carrier transport, and recombination dynamics within the dual-absorber structure.

### 4.5. Combined effect of CMTS and FeSi$_2$ doping density

The combined variation of CMTS and FeSi$_2$ acceptor doping densities revealed distinct trends in device performance (Fig. 7). A higher $V_{OC}$ was consistently achieved when the CMTS doping density was high (~$10^{21}$ cm$^{-3}$), regardless of the FeSi$_2$ doping level. In contrast, $J_{SC}$ was maximized at lower CMTS doping densities (<$10^{17}$ cm$^{-3}$). FF improved significantly at high CMTS doping levels (>$10^{20}$ cm$^{-3}$). The highest PCE was obtained when the CMTS doping was low (<$10^{17}$ cm$^{-3}$) and the FeSi$_2$ doping was high (>$10^{20}$ cm$^{-3}$), suggesting that the strong internal field in the heavily doped FeSi$_2$ layer promotes efficient hole transport toward the HTL and suppresses recombination at the interfaces, while the lightly doped CMTS layer ensures adequate minority-carrier diffusion length and favorable band alignment with the CdS ETL for efficient electron extraction. This balanced field distribution enables effective charge separation and an optimized trade-off among $V_{OC}$, $J_{SC}$, and FF.

### 4.6. ETL thickness and doping density

Following the analysis of absorber-layer parameters, the effect of the CdS ETL on device performance was examined by simultaneously varying its thickness and doping density while keeping other parameters constant (Fig. 8). The simulations show that $V_{OC}$ is largely insensitive to CdS doping at a fixed thickness, with only a minor decrease (~0.6 mV) at higher doping for thick CdS (> 0.7 μm), whereas at a fixed doping level, increasing CdS thickness causes a slight rise (~0.8 mV) in $V_{OC}$ for low doping ($10^{14}$–$10^{16}$ cm$^{-3}$), while at higher doping $V_{OC}$ becomes essentially independent of thickness. $J_{SC}$ remains nearly constant over the full range of CdS thicknesses and doping densities studied. FF shows minimal sensitivity overall as well; a modest improvement (up to ~1.5%) appears at higher doping when thickness > ~0.5 μm, whereas at a fixed doping, increasing CdS thickness leads to a slight decline (~0.5%) for very low doping levels ($10^{14}$–$10^{15}$ cm$^{-3}$). Consequently, PCE varies little across the explored CdS parameter space. A slight improvement (up to ~1.4%) is observed at higher doping levels when the ETL is thicker than ~0.5 μm, whereas at very low doping ($10^{14}$–$10^{15}$ cm$^{-3}$), increasing thickness leads to a modest decline in PCE.

These minimal sensitivities can be attributed to the intrinsic role of the CdS ETL as an electrical transport and buffer layer rather than an optically active component [82], [130], [131]. CdS is a wide-bandgap, optically transparent buffer that provides electron selectivity, hole blocking, and favorable band alignment with the absorber and TCO; it does not contribute to photogeneration. Once a minimum ETL thickness and doping (hence conductivity and interface quality) are achieved, the layer effectively supports the built-in potential and ensures selective carrier transport across the absorber–ETL interface, so further increases in thickness or doping only marginally affect field distribution, series resistance, or recombination.

### 4.7. HTL thickness and doping density

Similarly, after analyzing the ETL, the effect of the Cu$_2$O HTL on device performance was investigated by varying its thickness and doping density while keeping other parameters constant (Fig. 9). The simulation results show that variations in the HTL thickness and doping density have negligible influence on the key photovoltaic parameters of the device, consistent with the general trends reported in earlier studies [132]–[134]. At a fixed Cu$_2$O thickness, $V_{OC}$ increases only marginally (~0.2%) with higher Cu$_2$O doping, whereas at a fixed doping level, $V_{OC}$ remains nearly independent of the Cu$_2$O thickness. $J_{SC}$ remains nearly constant across all investigated ranges, with only a minimal rise (~0.09%) at higher thicknesses, whereas FF shows a minor enhancement (~0.5%) at moderate doping levels ($10^{17}$-$10^{19}$ cm$^{-3}$). Consequently, the PCE shows only a slight improvement (up to ~1.2%) with increasing Cu$_2$O doping, primarily reflecting the modest enhancements in $V_{OC}$ and FF, while remaining nearly insensitive (up to ~0.5% decrease at higher thicknesses) to Cu$_2$O thickness at a fixed doping level.

These weak dependencies arise because the Cu$_2$O HTL does not participate directly in photon absorption or carrier generation. Instead, it primarily ensures proper interfacial energetics by facilitating hole transport, blocking electron backflow, and reducing recombination at the FeSi$_2$/HTL interface. Once sufficient conductivity and interface quality are established, further variation in HTL thickness or doping has little impact on carrier transport or device performance.

### 4.8. CMTS and FeSi$_2$ bulk defect density

In thin-film solar cells, defects can arise both within the bulk absorber layers and at their interfaces. Bulk defects typically originate from intrinsic lattice imperfections and compositional non-stoichiometry, while interface defects often stem from lattice mismatch, or atomic interdiffusion between adjacent layers, which generate localized trap states that facilitate recombination [135][136]. The analysis in this section focuses on how variations in bulk defect densities in the CMTS and FeSi$_2$ absorbers affect overall device performance (Fig. 10). The results reveal distinct sensitivities of the two layers to defect-induced recombination, consistent with previously reported dual-absorber simulations [57], [123]. For a fixed CMTS defect density, $V_{OC}$ decreases sharply with increasing FeSi$_2$ defect density, whereas for a fixed FeSi$_2$ defect density, $V_{OC}$ shows only a modest decline with increasing CMTS defects. $J_{SC}$, in contrast, exhibits the opposite trend. Increasing CMTS defect density causes a pronounced reduction in $J_{SC}$ at any given FeSi$_2$ defect level, while changes in FeSi$_2$ defect density have only a minor effect on $J_{SC}$ when CMTS defects are fixed. The dependence of FF on defect density is more complex and nonlinear. When CMTS defects are held constant, FF initially increases slightly with FeSi$_2$ defect density up to about $10^{17}$ cm$^{-3}$, followed by a gradual decline at higher values. Conversely, for fixed FeSi$_2$ defect density, FF decreases with increasing CMTS defect density, showing a minor recovery at very high defect levels. However, such recovery likely arises from numerical artefacts or internal field redistribution and remains detrimental to charge extraction. PCE follows the combined trends of these parameters. For both cases, performance remains nearly stable up to moderate defect densities (~$10^{15}$ cm$^{-3}$) but declines sharply once this threshold is exceeded, driven by the concurrent reductions in $V_{OC}$, $J_{SC}$, and FF.

These trends can be explained by the carrier transport pathways and optical roles of the two absorbers. FeSi$_2$ defects have a stronger influence on $V_{OC}$ because the overall quasi-Fermi level splitting in a dual-absorber structure is primarily limited by the narrower bandgap absorber [123]. Increased recombination within FeSi$_2$ reduces its internal quasi-Fermi-level separation, thereby limiting the maximum attainable $V_{OC}$ of the entire device. The observed dependence of $J_{SC}$ suggests that carrier generation and collection are more strongly governed by the CMTS layer. This behavior can be attributed to CMTS absorbing a substantial portion of incident photons in the visible region, where the solar spectrum is most intense [103], while FeSi$_2$ contributes mainly at longer wavelengths. Consequently, defects in CMTS are more likely to influence the total photocurrent by introducing recombination centers that reduce carrier diffusion length and lifetime, resulting in lower collection efficiency [137]. The complex behavior of FF reflects the interplay between carrier transport and recombination losses within the multilayer structure, as also observed in other multi-absorber simulations [138]. While moderate defect densities may slightly modify internal electric fields in a way that temporarily improves charge selectivity, excessive defects inevitably introduce resistive and recombination losses that hinder efficient carrier extraction. Consequently, PCE integrates these effects: low-to-moderate defect densities (up to ~$10^{15}$ cm$^{-3}$) allow stable operation, whereas excessive defects in CMTS or FeSi$_2$ lead to lowering efficiency.

### 4.9. Interface defect density

Following the analysis of bulk defect densities in the CMTS and FeSi$_2$ absorbers, interfacial defect states were investigated to understand their impact on recombination dynamics and carrier transport (Fig. 11 and 12). Specifically, the defect densities at three key junctions: CMTS/FeSi$_2$ (primary/secondary absorber), CdS/CMTS (ETL/primary absorber), and FeSi$_2$/Cu$_2$O (HTL/secondary absorber) were systematically varied.

#### 4.9.1 CMTS/FeSi$_2$ interface

All the performance parameters show a consistent trend with respect to the interface defect density at the CMTS/FeSi$_2$ junction (Fig. 11). Up to a defect density of $10^{14}$ cm$^{-3}$, all four parameters remain nearly unchanged, indicating that the interface is relatively benign and does not introduce significant recombination or transport losses. This suggests that the defect states within this range are insufficient to introduce substantial non-radiative recombination losses. Beyond this threshold, however, a noticeable decline in all parameters is observed with increasing defect density, consistent with previous findings on interface-limited recombination in dual-absorber systems [57], [82], [91], [139]. This degradation can be attributed to enhanced recombination activity at the CMTS/FeSi$_2$ interface, where increasing defect density introduces additional recombination centers. As a result, the quasi-Fermi level splitting decreases, leading to a reduction in $V_{OC}$. Simultaneously, increased interfacial recombination shortens the minority carrier lifetime and reduces collection efficiency [137], thereby lowering $J_{SC}$. Moreover, the resulting difficulty in carrier extraction resembles the effects of higher series resistance, ultimately reducing the FF. The overall outcome is a decrease in PCE.

#### 4.9.2 CdS/CMTS and FeSi$_2$/Cu$_2$O interfaces

The CdS/CMTS interface (ETL/primary absorber junction) showed relatively minor sensitivity to increasing interface defect density (Fig. 12). Up to approximately $10^{15}$ cm$^{-3}$, $V_{oc}$, $J_{sc}$, FF, and PCE remained largely unaffected. This stability can be attributed to the efficient band alignment and minimal recombination at low defect levels, which allows for effective carrier extraction from the CMTS absorber. As the defect density increases beyond $10^{15}$ cm$^{-3}$, a gradual degradation in device performance is observed, with all parameters starting to decline. This trend suggests the onset of increased SRH recombination at the interface [140], hindering the separation and transport of photogenerated carriers. The performance saturation beyond ~$10^{17}$ cm$^{-3}$ indicates that the recombination rate reaches a limit, beyond which further increases in defect density have negligible additional effects on carrier dynamics. These behaviors align with general observations in other solar cell works, where interface recombination becomes dominant once defect density reaches certain thresholds [91], [141].

On the other hand, the FeSi$_2$/Cu$_2$O interface (secondary absorber/HTL junction) exhibits a stronger sensitivity (Fig. 12). $V_{OC}$ decreases continuously with increasing defect density, plateauing after ~$10^{17}$ cm$^{-3}$. This trend reflects enhanced recombination at the HTL interface, which reduces quasi-Fermi level splitting and consequently $V_{OC}$. $J_{sc}$ remains stable up to ~$10^{14}$ cm$^{-3}$ but declines thereafter, as recombination intensifies. FF exhibits a non-monotonic response: it increases slightly between $10^{10}$ and $10^{11}$ cm$^{-3}$ but decreases sharply up to $10^{16}$ cm$^{-3}$ as recombination and resistive losses dominate, followed by a slight recovery beyond $10^{16}$ cm$^{-3}$. The initial and later increase in FF can be attributed to the relatively unchanged maximum power point when either or both $V_{OC}$ and $J_{SC}$ decrease [116]. PCE decreases consistently with increasing defect density, showing trends consistent with other solar cell modelling studies on Cu$_2$O-based interfaces [116], [142].

**4.10. Front and back contact work functions**

This section examines how front and back contact work functions influence the performance parameters (Fig. 13 and 14). First, the effects of the front contact metal was investigated by varying the corresponding work function ($\Phi_{m\text{-front}}$) from 3.8 eV to 4.6 eV. The results show that $V_{OC}$ remains constant up to ~ 4.4 eV and decreases thereafter. $J_{SC}$ exhibits a slight decline at the lowest $\Phi_{m\text{-front}}$ values (around 3.8–3.9 eV) and then saturates across the higher range. Both FF and PCE remain stable up to ~ 4.4 eV before dropping sharply beyond this point.

These behaviors can be interpreted in terms of the energy alignment between the front metal and the ITO layer, which serves as the n-type TCO in the structure. ITO possesses an electron affinity of approximately 4.0 eV and a bandgap of ~3.5 eV, placing its conduction-band minimum near 4.0 eV below the vacuum level. When the metal work function is lower than or close to the ITO conduction-band level ($\Phi_{m\text{-front}} \leq$ ~4.4 eV) (Fig. 13b), the Fermi levels of the two materials align favorably, establishing an ohmic contact that facilitates efficient electron extraction from the device to the external circuit. This results in stable $V_{OC}$, FF, and PCE within this work function range.

However, as $\Phi_{m\text{-front}}$ exceeds ~ 4.4 eV, the energy alignment shifts such that the metal Fermi level moves deeper below the ITO conduction band, creating an electron injection barrier (Schottky barrier) at the interface. This barrier impedes electron transport from ITO to the front metal, increases contact resistance, and enhances interface recombination. Consequently, FF and PCE decrease sharply due to higher resistive losses and poorer carrier extraction efficiency. The decline in $V_{OC}$ at $\Phi_{m\text{-front}} >$ 4.4 eV is attributed to enhanced interfacial recombination and reduced quasi-Fermi level splitting in the absorber region, as part of the built-in potential is consumed in overcoming the contact barrier. The slight reduction in $J_{SC}$ at low $\Phi_{m\text{-front}}$ values (below 4.0 eV) may arise from subtle barrier formation or band bending at the ITO/metal interface, which slightly affects carrier collection but stabilizes as the contact becomes more ohmic.

Overall, these findings emphasize that achieving a front metal work function close to—but not exceeding—the ITO conduction-band level (~4.0–4.4 eV) is critical for maintaining efficient electron extraction and minimizing parasitic resistance. Similar front-contact-dependent performance behavior has been observed in other heterojunction solar cells, where improper work function alignment leads to Schottky barrier formation and degraded device metrics [143]–[148].

Besides, the performance of the simulated device was also analyzed as a function of the back contact work function ($\Phi_{m\text{-back}}$), varied from 4.9 eV to 5.65 eV. The results show that $V_{OC}$ initially decreases slightly up to ~ 5.0 eV and then stabilizes, while $J_{SC}$ remains almost constant up to 5.5 eV, exhibiting a slight increase thereafter. Both FF and PCE increase sharply up to ~ 5.0 eV and then reach saturation, indicating the formation of an optimal contact regime.

These behaviors can be understood from the band alignment between the back metal and the adjacent $Cu_2O$ HTL. $Cu_2O$ is a p-type semiconductor with an electron affinity of ~ 3.2 eV and a bandgap of ~2.2 eV, placing its valence-band maximum around 5.4 eV below the vacuum level, defining its ionization energy ($\Phi_{IE}$). When the metal work function is significantly lower than this ($\Phi_{m\text{-back}} <$ 5.0 eV) (Fig. 14b), a Schottky barrier forms at the $Cu_2O$/metal interface, impeding hole extraction, increasing contact resistance, and promoting interfacial recombination—thereby reducing FF and PCE. As $\Phi_{m\text{-back}}$ approaches 5.0 eV, the barrier height decreases and the contact becomes increasingly ohmic, facilitating hole transport and improving device performance. Beyond ~5.0 eV, the Fermi level of the metal aligns closely with the $Cu_2O$ valence band, producing nearly ideal hole extraction with little further improvement, hence the observed saturation in FF and PCE. $J_{SC}$ remains nearly constant over most of the $\Phi_{m\text{-back}}$ range since photogeneration within the CMTS and $FeSi_2$ absorbers dominates over interfacial limitations, with only a marginal rise at $\Phi_{m\text{-back}} \geq 5.5$ eV due to improved hole extraction and reduced interfacial recombination.

These trends are consistent with earlier reports showing that tuning the metal work function near the valence-band edge of HTL converts the back contact from Schottky to ohmic, thereby improving carrier extraction and overall device performance [110], [141], [149], [150].

**4.11 Series and shunt resistance**

Series resistance ($R_{series}$) and shunt resistance ($R_{shunt}$) are critical parasitic parameters that influence carrier transport and overall device efficiency in solar cells (Fig. 15). $R_{series}$ represents the total resistive losses encountered by charge carriers during transport and extraction, originating primarily from the bulk resistances of the absorber and transport layers, the ETL/absorber and HTL/absorber interfaces, and the contact resistance at the metal electrodes [103], [116], [148], [151]. On the other hand, $R_{shunt}$ arises from unintended leakage pathways across the junction, typically introduced by pinholes, interfacial imperfections, or process-induced defects, which allow photogenerated carriers to recombine through alternate current paths [103]. Ideally, a solar cell should possess a very low $R_{series}$ and a very high $R_{shunt}$ to ensure efficient carrier transport and minimal parasitic losses.

Our simulated results (Fig. 16a-d) indicate that $V_{OC}$ remains nearly constant across a wide range of $R_{series}$ values. $J_{SC}$ also shows minimal sensitivity to $R_{series}$ variations. However, FF decreases steadily with increasing $R_{series}$, leading to a corresponding drop in PCE [73], [116], [148]. These trends can be explained by examining the role of $R_{series}$ under different operating conditions. Since $V_{OC}$ is determined by the quasi-Fermi level separation under open-circuit conditions, it is largely unaffected by $R_{series}$ because no current flows through resistive components (Fig. 15). $J_{SC}$ remains largely unaffected by $R_{series}$ since, under short-circuit conditions, the external voltage is zero and current collection is driven predominantly by the built-in electric field. Series resistance only becomes influential under load, where voltage drops occur during carrier transport. On the other hand, FF and PCE degrade as $R_{series}$ increases because higher resistive losses reduce the terminal voltage under current flow and lower the maximum power output (Fig. 17a)

Regarding the $R_{shunt}$ analysis, the simulation results show that $V_{OC}$ and FF are strongly dependent on $R_{shunt}$, while $J_{SC}$ remains comparatively stable across the examined range [133], [142] (Fig. 16f-h). As $R_{shunt}$ decreases below $10^2$ $\Omega \cdot cm^2$, $V_{OC}$ drops sharply because low shunt resistance introduces leakage paths that bypass the p–n junction (Fig. 15), reducing the voltage buildup under open-circuit conditions. Similarly, FF declines markedly below $10^3$ $\Omega \cdot cm^2$, as these leakage currents make the J–V curve less rectangular near the maximum power point (Fig. 17b) and reduce the usable output power. In contrast, when $R_{shunt}$ is sufficiently high ($\geq 10^3$ $\Omega \cdot cm^2$), both $V_{OC}$ and FF stabilize, resulting in a saturated PCE plateau, since minimal current leakage preserves ideal diode behavior. The negligible variation of $J_{SC}$ confirms that short-circuit current collection is governed mainly by the built-in electric field, and largely unaffected by shunt-related leakage paths. These results highlight the crucial role of $R_{shunt}$ in maintaining photovoltaic performance—low $R_{shunt}$ leads to parasitic current losses, while high $R_{shunt}$ ensures efficient carrier extraction and voltage retention. Achieving high $R_{shunt}$ through improved film uniformity, defect passivation, and mitigation of pinholes or grain-boundary pathways is therefore essential for sustaining high $V_{OC}$, FF, and overall device efficiency [116], [152]–[155].

**4.12 Temperature**

The thermal stability of the device was assessed by analyzing the influence of temperature variation on its photovoltaic parameters (Fig. 18). $V_{OC}$ showed a monotonic decrease with increasing temperature over the simulated range of 275 K to 500 K. This behavior arises from thermally induced modifications of semiconductor properties. The bandgap of most semiconductors narrows with the rise in temperature [156]–[158], which increases the intrinsic carrier concentration, and consequently enhances the reverse saturation current density ($J_0$). This relationship can be expressed as [159]:

$$J_0 \propto T^3 exp\left(\frac{-E_g}{kT}\right) \quad (10)$$

The $T^3$ dependence originates from the temperature scaling of the effective density of states in the conduction and valence bands [73], [94], [159], while the exponential term captures the sensitivity to bandgap narrowing. The increase in $J_0$ leads to a logarithmic decrease in $V_{oc}$ with temperature, as described by equation (7) [159].

$J_{SC}$ remained nearly unchanged with increasing temperature. While bandgap narrowing at elevated temperatures could theoretically improve low-energy photon absorption, the impact on overall carrier generation is minimal compared to the significant increase in recombination. Thermal excitation activates defect states and enhances non-radiative recombination in the bulk and at interfaces [156], effectively counterbalancing any gains in photogeneration. As a result, no significant change in $J_{SC}$ was observed, indicating a balance between generation and recombination processes in the structure.

FF showed a consistent decrease with temperature, which can be attributed to increased resistive and recombination losses. Higher temperatures degrade carrier mobility, increasing series resistance, and activate additional trap states, leading to enhanced recombination. Additionally, the reduction in $V_{OC}$ impacts the quality of the diode behavior, reducing the steepness of the J–V curve near the maximum power point. These factors collectively lower the FF. As a result, PCE decreased steadily with increasing temperature (showing a temperature coefficient of -0.049% $K^{-1}$), following the combined trends in $V_{OC}$ and FF. Similar temperature-dependent variations in $V_{OC}$, $J_{SC}$, FF, and PCE have also been reported in earlier works on solar cells [72], [73], [160].

### 4.13 Incident light intensity

The variation of photovoltaic parameters with incident light intensity was explored to gain insight into the device's photoresponse characteristics (Fig. 19). $V_{OC}$ shows a logarithmic increase with rising light intensity, consistent with the theoretical expression as given in equation (7). Since $J_{SC}$ increases linearly with illumination while $J_0$ remains nearly constant, $V_{OC}$ exhibits a logarithmic dependence. Physically, higher illumination generates more photocarriers, enhancing the quasi-Fermi level splitting within the absorber layers and thus increasing $V_{OC}$. $J_{SC}$ increases linearly with incident light intensity, as the number of photogenerated electron–hole pairs is directly proportional to the photon flux [161]. FF, however, shows a gradual decline with increasing light intensity. At higher current densities, the influence of series resistance becomes more pronounced, leading to voltage drops and reducing the squareness of the J–V curve. Additionally, the activation of localized defect states under intense illumination may further contribute to the observed FF reduction. Despite this and the increased input power, PCE increases modestly with light intensity, as the concurrent enhancements in $V_{OC}$ and $J_{SC}$ outweigh the opposing effects, resulting in a net rise in conversion efficiency [162]. Moreover, the device sustains a notable efficiency of 25.3% even at a low illumination of 200 W/m$^2$, underscoring its excellent weak-light response and strong potential for indoor photovoltaic applications.

### 4.14 Optimized device

Based on the preceding analyses, the material and device parameters were adjusted to achieve optimal photovoltaic performance at standard temperature and illumination (300 K and 1000 W/m$^2$ respectively). The optimized device delivers $V_{OC}$ = 0.79V, $J_{SC}$ = 51.07 mA/cm$^2$, FF = 85.91%, and PCE = 34.9%. The corresponding J–V and QE responses are shown in Fig. 20. The results were further compared with previously reported CMTS-based solar cells from both experimental and simulation studies (Table 4). Reported PCE values in the literature range from 0.14% to 31.51%, all lower than that achieved by the optimized structure in this work. These outcomes underscore the potential of the CMTS/FeSi$_2$ configuration as an efficient and practical solar cell architecture.

Table 4. Comparison of photovoltaic performance between the proposed structure and literature-reported solar cell designs

| Device structure | $V_{OC}$ (V) | $J_{SC}$ (mA/cm$^2$) | FF (%) | PCE (%) | Reference | Year |
|---|---|---|---|---|---|---|
| **Experimental works** | | | | | | |
| AZO/iZnO/CdS/CMTS/Mo/glass | 0.3084 | 4.7 | 33.9 | 0.49 | [163] | 2015 |
| Ag/AZO/i-ZnO/CdS/CMTS/Mo/SLG | 0.381 | 4.95 | 38.64 | 0.73 | [37] | 2016 |
| AZO/i-ZnO/CdS/CMTS/Mo/glass | 0.359 | 2.95 | 35.8 | 0.38 | [38] | 2016 |
| Ni/Al/AZO/i-ZnO/CdS/CMTS/Mo/Glass | 0.315 | 5.3 | 37 | 0.7 | [49] | 2016 |
| i-ZnO+AZO/CdS/CMTS/Mo | 0.354 | 5.8 | 40 | 0.83 | [48] | 2017 |
| Al/AZO/i-ZnO/CdS/CMTS/Mo | 0.226 | 4 | 36.3 | 0.33 | [44] | 2017 |
| Al/AZO/i-ZnO/CdS/CMTS/CdS/MoS$_2$/Mo | 0.289 | 1.59 | 29.9 | 0.14 | [164] | 2018 |
| Ag/AZO/i-ZnO/CdS/CMTS/FTO/glass | 0.372 | 6.82 | 31 | 0.76 | [34] | 2018 |
| Al/ITO/i-ZnO/CdS/CMTS/Mo/glass | 0.157 | 16.99 | 29.2 | 0.78 | [50] | 2019 |
| Al/AZO/iZnO/CdS/CMTS/Mo | 0.445 | 7.45 | 34 | 1.13 | [47] | 2023 |
| **Simulation works** | | | | | | |
| ZnO/CdS/CMTS/Back contact | 0.88 | 24.1 | 77.9 | 16.5 | [8] | 2018 |
| AZO/i-ZnO/CdS/CMTS/Back contact | 1.11 | 26.26 | 61.08 | 17.81 | [56] | 2022 |
| AZO/i-ZnO/Zn(O, S)/CMTS/Back contact | 1.11 | 26.27 | 66.22 | 19.45 | [56] | 2022 |
| AZO/i-ZnO/SnS$_2$/CMTS/Back contact | 1.12 | 26.44 | 68.33 | 20.26 | [56] | 2022 |
| ZnO:Al/ZnMgO/ZnSnO/CMTS | 1.107 | 23.49 | 79.38 | 20.66 | [4] | 2023 |
| ZnO:Al/ZnMgO/CdS/CMTS/Back contact | 1.096 | 23.47 | 81.32 | 20.92 | [4] | 2023 |
| ZnO:Al/ZnMgO/ZnSe/CMTS | 1.221 | 22.99 | 84.35 | 23.69 | [4] | 2023 |
| ZnO:Al/ZnMgO/Zn(O, S)/CMTS | 1.223 | 23.63 | 84.67 | 24.46 | [4] | 2023 |
| ZnO:Al/ZnMgO/SnS$_2$/CMTS | 1.236 | 23.69 | 85.49 | 25.03 | [4] | 2023 |
| Cu/ZnO:Al/i-ZnO/n-CdS/p-CMTS/Pt | 0.883 | 34.41 | 83.74 | 25.43 | [7] | 2023 |
| Cu/ZnO:Al/i-ZnO/n-CdS/p-CMTS/p+-SnS/Pt | 1.074 | 36.21 | 81.04 | 31.51 | [7] | 2023 |
| Al/AZO/i-ZnO/CdS/CMTS/Cu$_2$O/Mo | 1.26 | 24.45 | 70.85 | 21.78 | [10] | 2023 |
| Al/ITO/WS$_2$/CMTS/SnS/Mo | 1.1504 | 22.74 | 74.3 | 19.43 | [16] | 2025 |
| Al/ITO/SnS$_2$/CMTS/CZTS/Mo | 1.3123 | 23.95 | 88.16 | 27.71 | [16] | 2025 |
| ITO/TiO$_2$/CMTS/Sb$_2$Se$_3$/Mo | 0.78 | 25.42 | 65 | 12.86 | [2] | 2025 |
| Al/ITO/i-ZnO/CdS/CMTS/V$_2$O$_5$/Au | 1.02 | 28.7 | 85.89 | 25.2 | [165] | 2025 |
| Al/ITO/CdS/CMTS/FeSi$_2$/Cu$_2$O/Au | 0.79 | 51.07 | 85.91 | 34.9 | **This work** | 2025 |

## 5. Conclusion

This work presents a comprehensive computational investigation of a novel CMTS/FeSi$_2$-based dual-absorber thin-film solar cell with the configuration Al/ITO/CdS/CMTS/FeSi$_2$/Cu$_2$O/Au, aiming to enhance photovoltaic performance through careful tuning of material and device parameters. The inclusion of FeSi$_2$ as a secondary absorber beneath CMTS notably enhanced light harvesting, as evidenced by the substantial increase in J$_{SC}$ from 19.31 to 51.04 mA cm$^{-2}$, despite a moderate reduction in V$_{OC}$. This extended photoresponse, confirmed by the broadened quantum efficiency spectrum, demonstrates the synergistic role of the FeSi$_2$ layer in capturing near-infrared photons that are otherwise unabsorbed in single-absorber CMTS devices. Subsequent analyses identified the interplay between absorber layers' thickness and doping density as critical to optimizing carrier transport and recombination dynamics. The results revealed that a lightly doped and thinner CMTS layer (<10$^{17}$ cm$^{-3}$, <0.2 μm) combined with a thicker, heavily doped FeSi$_2$ layer (>0.5 μm, >10$^{20}$ cm$^{-3}$) yields the best performance balance, promoting efficient charge separation and extraction. The ETL and HTL studies further indicated that variations in their thickness and doping have only minor effects on device performance. Defect density analyses showed that device performance is highly sensitive to recombination in both bulk and interfacial regions. Bulk defects in the absorbers above ~10$^{15}$ cm$^{-3}$ significantly degraded the performance parameters, with CMTS defects having stronger effects on current density and FeSi$_2$ defects more strongly influencing voltage. Among the interfaces, the CMTS/FeSi$_2$ junction exhibited a pronounced sensitivity to defect density exceeding 10$^{14}$ cm$^{-3}$, beyond which the device performance deteriorated significantly. The ETL/CMTS interface maintained stable operation up to moderate defect levels (~10$^{15}$ cm$^{-3}$), with only modest degradation observed thereafter. The FeSi$_2$/HTL interface, however, showed a steady decline in efficiency as defect density increased. These findings highlight the critical importance of effective interface passivation during fabrication to mitigate recombination losses and preserve device performance. Contact and parasitic resistance parameter studies demonstrated that while the front and back metal work functions influence band alignment and charge extraction, the series and shunt resistances play a greater role in determining FF and PCE. Thermal and illumination analyses further showed that PCE decreases with increasing temperature (with a temperature coefficient of −0.049% K$^{−1}$), whereas higher light intensity slightly improves efficiency due to enhanced carrier generation. Upon optimizing all critical parameters, the proposed dual-absorber structure achieved a high V$_{OC}$ = 0.79 V, J$_{SC}$ = 51.07 mA cm$^{−2}$, FF = 85.91%, and PCE = 34.9% under standard conditions (300 K, 1000 Wm$^{−2}$). Reported PCEs for similar single-absorber CMTS solar cells range from 0.14% to 31.51%, confirming the advancement achieved in this work. Beyond the simulated results, the constituent materials are largely earth-abundant, chemically stable, and compatible with established thin-film deposition techniques, making the proposed structure experimentally realizable. Each layer — from the CMTS and FeSi$_2$ absorbers to the CdS buffer and Cu$_2$O HTL — can be fabricated using scalable routes such as sputtering, spray pyrolysis, electrodeposition, or chemical bath deposition. The present findings thus provide a strong theoretical foundation for experimental exploration of CMTS/FeSi$_2$-based solar cells, offering clear guidance on optimal doping levels, defect tolerances, and interfacial requirements. While this study provides important design insights, it remains computational in nature. Future work should focus on experimentally fabricating and characterizing the proposed structure to validate its optoelectronic behavior under real operating conditions. Investigations into interface passivation strategies, graded junctions, and long-term stability will be crucial to translating these findings into practical, high-efficiency thin-film photovoltaics. Overall, this work demonstrates the significant potential of CMTS/FeSi$_2$ dual-absorber solar cells as an efficient, scalable, and sustainable pathway toward next-generation photovoltaic technologies.

References

[1] A. Hammoud *et al.*, "Investigation on Cu2MgSnS4 thin film prepared by spray pyrolysis for photovoltaic and humidity sensor applications," *Opt. Mater. (Amst).*, vol. 127, p. 112296, May 2022, doi: 10.1016/j.optmat.2022.112296.

[2] S. S. Bal, A. Basak, and U. P. Singh, "Performance enhancement of Sb$_2$Se$_3$ solar cells via bilayer architecture: a SCAPS-1D simulation study," *J. Mater. Sci. Mater. Electron.*, vol. 36, no. 14, p. 811, May 2025, doi: 10.1007/s10854-025-14857-1.

[3] H. Tong *et al.*, "Total-area world-record efficiency of 27.03% for 350.0 cm2 commercial-sized single-junction silicon solar cells," *Nat. Commun.*, vol. 16, no. 1, p. 5920, Jul. 2025, doi: 10.1038/s41467-025-61128-y.

[4] A. Kowsar *et al.*, "Enhanced photoconversion efficiency of Cu2MnSnS4 solar cells by Sn-/Zn-based oxides and chalcogenides buffer and electron transport layers," *Sol. Energy*, vol. 265, no. July, p. 112096, 2023, doi: 10.1016/j.solener.2023.112096.

[5] H. Lin *et al.*, "Silicon heterojunction solar cells with up to 26.81% efficiency achieved by electrically optimized nanocrystalline-silicon hole contact layers," *Nat. Energy*, vol. 8, no. 8, pp. 789–799, 2023, doi: 10.1038/s41560-023-01255-2.

[6] W. Shockley and H. J. Queisser, "Detailed balance limit of efficiency of p-n junction solar cells," *J. Appl. Phys.*, vol. 32, no. 3, pp. 510–519, 1961, doi: 10.1063/1.1736034.

[7] A. Isha *et al.*, "High efficiency Cu2MnSnS4 thin film solar cells with SnS BSF and CdS ETL layers: A numerical simulation," *Heliyon*, vol. 9, no. 5, p. e15716, 2023, doi: 10.1016/j.heliyon.2023.e15716.

[8] Y. H. Khattak, F. Baig, B. Marí, S. Beg, and S. Ahmed, " Baseline for the Numerical Analysis of High Efficiency Copper Manganese Tin Sulfide Cu 2 MnSnS 4 Based Thin Film Solar Cell ," *J. Nanoelectron. Optoelectron.*, vol. 13, no. 11, pp. 1678–1684, 2018, doi: 10.1166/jno.2018.2421.

[9] S. Moujoud *et al.*, "Performance analysis of CuInSe2 based solar cells using SCAPS-1D," *Mater. Today Proc.*, vol. 66, pp. 17–21, 2022, doi: 10.1016/j.matpr.2022.03.101.

[10] W. Henni *et al.*, "Effect of Adding Cu2O as a Back Surface Field Layer on the Performance of Copper Manganese Tin Sulfide Solar Cells," *Sustainability*, vol. 15, no. 19, p. 14322, Sep. 2023, doi: 10.3390/su151914322.

[11] R. Kumar, A. Kumar, R. Pushkar, and A. Priyadarshi, "The impact of SnMnO2 TCO and Cu2O as a hole transport layer on CIGSSe solar cell performance improvement," *J. Comput. Electron.*, vol. 22, no. 4, pp. 1107–1127, 2023, doi: 10.1007/s10825-023-02050-8.

[12] A. Hammoud, B. Yahmadi, M. Souli, S. A. Ahmed, L. Ajili, and N. Kamoun-Turki, "Effect of sulfur content on improving physical properties of new sprayed Cu2MgSnS4 thin films compound for optoelectronic applications," *Eur. Phys. J. Plus*, vol. 137, no. 2, p. 232, Feb. 2022, doi: 10.1140/epjp/s13360-022-02417-z.

[13] J. Keller *et al.*, "High-concentration silver alloying and steep back-contact gallium grading enabling copper indium gallium selenide solar cell with 23.6% efficiency," *Nat. Energy*, vol. 9, no. 4, pp. 467–478, 2024, doi: 10.1038/s41560-024-01472-3.

[14] M. A. Green *et al.*, "Solar Cell Efficiency Tables (Version 66)," *Prog. Photovoltaics Res. Appl.*, vol. 33, no. 7, pp. 795–810, Jul. 2025, doi: 10.1002/pip.3919.

[15] M. Nakamura, K. Yamaguchi, Y. Kimoto, Y. Yasaki, T. Kato, and H. Sugimoto, "Cd-Free Cu(In,Ga)(Se,S)2 thin-film solar cell with record efficiency of 23.35%," *IEEE J. Photovoltaics*, vol. 9, no. 6, pp. 1863–1867, 2019, doi: 10.1109/JPHOTOV.2019.2937218.

[16] N. Mahsar, B. Zaidi, L. Dehimi, A. Barkhordari, and F. Hadef, "The performance of Cu2MnSnS4 based solar cells with CZTS as hole transport nanolayer," *Sci. Rep.*, vol. 15, no. 1, p. 26774, Jul. 2025, doi: 10.1038/s41598-025-12766-1.

[17] A. E. H. Benzetta, M. Abderrezek, and M. E. Djeghlal, "Contribution to improve the performances of Cu2ZnSnS4 thin-film solar cell via a back surface field layer," *Optik (Stuttg).*, vol. 181, pp. 220–230, 2019, doi:

[18] J. M. Burst *et al.*, "CdTe solar cells with open-circuit voltage breaking the 1V barrier," *Nat. Energy*, vol. 1, no. 4, 2016, doi: 10.1038/NENERGY.2016.15.

[19] S. Rao, A. Morankar, H. Verma, and P. Goswami, "Emerging Photovoltaics: Organic, Copper Zinc Tin Sulphide, and Perovskite-Based Solar Cells," *J. Appl. Chem.*, vol. 2016, pp. 1–12, 2016, doi: 10.1155/2016/3971579.

[20] A. Sharma, P. Sahoo, A. Singha, S. Padhan, G. Udayabhanu, and R. Thangavel, "Efficient visible-light-driven water splitting performance of sulfidation-free, solution processed Cu2MgSnS4 thin films: Role of post-drying temperature," *Sol. Energy*, vol. 203, pp. 284–295, Jun. 2020, doi: 10.1016/j.solener.2020.04.027.

[21] C. T. Illiyas and K. C. Preetha, "An Investigation on Structural, Morphological, Optical, and Electrical Properties of Copper Zinc Tin Sulfide (CZTS) Thin Films Prepared by SILAR Method," *Brazilian J. Phys.*, vol. 54, no. 4, pp. 1–13, 2024, doi: 10.1007/s13538-024-01495-x.

[22] N. Mekhaznia and B. Zaidi, "Detailed performance analysis of FTO/TiO2/FAPbI3/CZTS solar cells: computational study," vol. 13401, no. Icait, p. 24, 2024, doi: 10.1117/12.3052232.

[23] B. Zaidi, C. Shekhar, K. Kamli, Z. Hadef, S. Belghit, and M. S. Ullah, "Junction configuration effect on the performance of In2S3/CZTS solar cells," *J. Nano- Electron. Phys.*, vol. 12, no. 1, pp. 3–5, 2020, doi: 10.21272/jnep.12(1).01024.

[24] I. M. El Radaf, "Intensive studies on the structural, optical, and optoelectrical properties of a novel kesterite Cu CdGeS4 thin films," *Optik (Stuttg).*, vol. 272, p. 170358, Feb. 2023, doi: 10.1016/j.ijleo.2022.170358.

[25] H. Katagiri *et al.*, "Development of CZTS-based thin film solar cells," *Thin Solid Films*, vol. 517, no. 7, pp. 2455–2460, Feb. 2009, doi: 10.1016/j.tsf.2008.11.002.

[26] M. A. Green *et al.*, "Solar cell efficiency tables (Version 61)," *Prog. Photovoltaics Res. Appl.*, vol. 31, no. 1, pp. 3–16, Jan. 2023, doi: 10.1002/pip.3646.

[27] S. Sun *et al.*, "Influence of Cd0.6Zn0.4S buffer layer on the band alignment and the performance of CZTS thin film solar cells," *Opt. Mater. (Amst).*, vol. 112, p. 110666, Feb. 2021, doi: 10.1016/j.optmat.2020.110666.

[28] M. S. Kumar, S. P. Madhusudanan, and S. K. Batabyal, "Substitution of Zn in Earth-Abundant Cu2ZnSn(S,Se)4 based thin film solar cells – A status review," *Sol. Energy Mater. Sol. Cells*, vol. 185, pp. 287–299, Oct. 2018, doi: 10.1016/j.solmat.2018.05.003.

[29] B. S. Pawar *et al.*, "Effect of complexing agent on the properties of electrochemically deposited Cu2ZnSnS4 (CZTS) thin films," *Appl. Surf. Sci.*, vol. 257, no. 5, pp. 1786–1791, Dec. 2010, doi: 10.1016/j.apsusc.2010.09.016.

[30] K. Hartman *et al.*, "Detection of ZnS phases in CZTS thin-films by EXAFS," *Conf. Rec. IEEE Photovolt. Spec. Conf.*, pp. 002506–002509, 2011, doi: 10.1109/PVSC.2011.6186455.

[31] K. Yin *et al.*, "Gradient bandgaps in sulfide kesterite solar cells enable over 13% certified efficiency," *Nat. Energy*, vol. 10, no. 2, pp. 205–214, 2025, doi: 10.1038/s41560-024-01681-w.

[32] S. Lie, M. Guc, V. Tunuguntla, V. Izquierdo-Roca, S. Siebentritt, and L. H. Wong, "Comprehensive physicochemical and photovoltaic analysis of different Zn substitutes (Mn, Mg, Fe, Ni, Co, Ba, Sr) in CZTS-inspired thin film solar cells," *J. Mater. Chem. A*, vol. 10, no. 16, pp. 9137–9149, 2022, doi: 10.1039/D2TA00225F.

[33] H. Mebrek, B. Zaidi, N. Mekhaznia, H. Al-Dmour, and A. Barkhordari, "Performance evaluation of Cu2SrSnS4 based solar cell: effect of transition metal dichalcogenides buffer layer," *Sci. Rep.*, vol. 15, no. 1, p. 6694, Feb. 2025, doi: 10.1038/s41598-025-91145-2.

[34] J. Yu *et al.*, "Improvement performance of two-step electrodepositing Cu 2 MnSnS 4 thin film solar cells by tuning Cu-Sn alloy layer deposition time," *Mater. Chem. Phys.*, vol. 211, pp. 382–388, Jun. 2018, doi: 10.1016/j.matchemphys.2018.03.009.

[35] M. A. Green, E. D. Dunlop, D. H. Levi, J. Hohl-Ebinger, M. Yoshita, and A. W. Y. Ho-Baillie, "Solar cell efficiency tables (version 54)," *Prog. Photovoltaics Res. Appl.*, vol. 27, no. 7, pp. 565–575, Jul. 2019, doi: 10.1002/pip.3171.

[36] J. Jean, P. R. Brown, R. L. Jaffe, T. Buonassisi, and V. Bulović, "Pathways for solar photovoltaics," *Energy Environ. Sci.*, vol. 8, no. 4, pp. 1200–1219, 2015, doi: 10.1039/C4EE04073B.

[37] R. R. Prabhakar *et al.*, "Photovoltaic effect in earth abundant solution processed Cu2MnSnS4 and Cu2MnSn(S,Se)4 thin films," *Sol. Energy Mater. Sol. Cells*, vol. 157, pp. 867–873, 2016, doi:

[38] L. Chen *et al.*, "Strategic improvement of Cu2MnSnS4 films by two distinct post-annealing processes for constructing thin film solar cells," *Acta Mater.*, vol. 109, pp. 1–7, 2016, doi: 10.1016/j.actamat.2016.02.057.

[39] A. Ziti *et al.*, "Growth and characterization of pure stannite Cu2MnSnS4 thin films deposited by dip-coating technique," *Appl. Phys. A*, vol. 127, no. 9, p. 663, Sep. 2021, doi: 10.1007/s00339-021-04824-y.

[40] A. Bouali, O. Kamoun, M. Hajji, I. N. Popescu, R. Vidu, and N. Turki Kamoun, "Improving CMTS Physical Properties Through Potassium Doping for Enhanced Rhodamine B Degradation," *Technologies*, vol. 13, no. 7, pp. 1–19, 2025, doi: 10.3390/technologies13070301.

[41] M. A. Chenafi, A. Boucif, D. E. Mellah, and K. Demmouche, "First-principles calculation of structural and optoelectronic properties of Cu2MgSnS4 (CMTS): critical insights from meta-GGA," *Phys. Scr.*, vol. 100, no. 7, 2025, doi: 10.1088/1402-4896/addf93.

[42] H. Guan, H. Hou, M. Li, and J. Cui, "Photocatalytic and thermoelectric properties of Cu 2 MnSnS 4 nanoparticles synthesized via solvothermal method," *Mater. Lett.*, vol. 188, pp. 319–322, Feb. 2017, doi: 10.1016/j.matlet.2016.09.018.

[43] S. Lie *et al.*, "Improving the charge separation and collection at the buffer/absorber interface by double-layered Mn-substituted CZTS," *Sol. Energy Mater. Sol. Cells*, vol. 185, pp. 351–358, Oct. 2018, doi: 10.1016/j.solmat.2018.05.052.

[44] S. Marchionna, A. Le Donne, M. Merlini, S. Binetti, M. Acciarri, and F. Cernuschi, "Growth of Cu2MnSnS4PV absorbers by sulfurization of evaporated precursors," *J. Alloys Compd.*, vol. 693, pp. 95–102, 2017, doi: 10.1016/j.jallcom.2016.09.176.

[45] K. Rudisch, W. F. Espinosa-García, J. M. Osorio-Guillén, C. M. Araujo, C. Platzer-Björkman, and J. J. S. Scragg, "Structural and Electronic Properties of Cu 2 MnSnS 4 from Experiment and First-Principles Calculations," *Phys. status solidi*, vol. 256, no. 7, Jul. 2019, doi: 10.1002/pssb.201800743.

[46] J. Yu *et al.*, "Synthesis of Cu2MnSnS4 thin film deposited on seeded fluorine doped tin oxide substrate via a green and low-cost electrodeposition method," *Mater. Lett.*, vol. 191, pp. 186–188, Mar. 2017, doi: 10.1016/j.matlet.2016.12.067.

[47] V. Trifiletti *et al.*, "Manganese-substituted kesterite thin-films for earth-abundant photovoltaic applications," *Sol. Energy Mater. Sol. Cells*, vol. 254, p. 112247, Jun. 2023, doi: 10.1016/j.solmat.2023.112247.

[48] A. Le Donne, S. Marchionna, M. Acciarri, F. Cernuschi, and S. Binetti, "Relevant efficiency enhancement of emerging Cu2MnSnS4 thin film solar cells by low temperature annealing," *Sol. Energy*, vol. 149, pp. 125–131, 2017, doi: 10.1016/j.solener.2017.03.087.

[49] B. Ananthoju, J. Mohapatra, M. K. Jangid, D. Bahadur, N. V. Medhekar, and M. Aslam, "Cation/Anion Substitution in Cu2ZnSnS4 for Improved Photovoltaic Performance," *Sci. Rep.*, vol. 6, no. September, pp. 1–11, 2016, doi: 10.1038/srep35369.

[50] G. Yang *et al.*, "Synthesis and characterizations of Cu 2 MgSnS 4 thin films with different sulfuration temperatures," *Mater. Lett.*, vol. 242, pp. 58–61, 2019, doi: 10.1016/j.matlet.2019.01.102.

[51] M. F. Rahman *et al.*, "Improving the efficiency of a CIGS solar cell to above 31% with Sb2S3 as a new BSF: a numerical simulation approach by SCAPS-1D," *RSC Adv.*, vol. 14, no. 3, pp. 1924–1938, 2024, doi: 10.1039/d3ra07893k.

[52] A. A. El-Naggar *et al.*, "SCAPS simulation and design of highly efficient CuBi2O4-based thin-film solar cells (TFSCs) with hole and electron transport layers," *Sci. Rep.*, vol. 15, no. 1, pp. 1–25, 2025, doi: 10.1038/s41598-025-12091-7.

[53] R. Uddin, S. U. Alam, and S. Bhowmik, "Comprehensive Analysis for Low-Cost and Highly Efficient Perovskite Solar Cells Using SCAPS-1D with an Inexpensive Hole Transport Material, Electron Transport Material, and Back Contact Considering the Toxicity," *ACS Omega*, vol. 10, no. 34, pp. 38480–38497, 2025, doi: 10.1021/acsomega.5c01202.

[54] M. A. Siddika, M. R. I. Sheikh, M. F. Ali, A. Al Mamun, and M. J. Hossen, "Enhanced efficiency of thin-film solar cells using AgInSe2 back surface field layer: a SCAPS-1D numerical study," *Mater. Technol.*, vol. 40, no. 1, pp. 1–15, 2025, doi: 10.1080/10667857.2025.2562530.

[55] L. A. Lotfy, M. Abdelfatah, S. W. Sharshir, A. A. El-Naggar, W. Ismail, and A. El-Shaer, "Numerical simulation and optimization of FTO/TiO2/CZTS/CuO/Au solar cell using SCAPS-1D," *Sci. Rep.*, vol. 15, no. 1, pp. 1–24, 2025, doi: 10.1038/s41598-025-12999-0.

[56] T. Pansuriya, R. Malani, and V. Kheraj, "Investigations on the effect of buffer layer on CMTS based thin film solar cell using SCAPS 1-D," *Opt. Mater. (Amst).*, vol. 126, no. March, p. 112150, 2022, doi: 10.1016/j.optmat.2022.112150.

[57] S. Hemalatha, R. T. Prabu, R. Radhika, and A. Kumar, "Dual-Absorber Thin-Film Solar Cell: A High-Efficiency Design," *Phys. Status Solidi Appl. Mater. Sci.*, vol. 220, no. 9, 2023, doi: 10.1002/pssa.202200761.

[58] F. H. Alharbi and S. Kais, "Theoretical limits of photovoltaics efficiency and possible improvements by intuitive approaches learned from photosynthesis and quantum coherence," *Renew. Sustain. Energy Rev.*, vol. 43, pp. 1073–1089, Mar. 2015, doi: 10.1016/j.rser.2014.11.101.

[59] S. Rühle, "Tabulated values of the Shockley–Queisser limit for single junction solar cells," *Sol. Energy*, vol. 130, pp. 139–147, Jun. 2016, doi: 10.1016/j.solener.2016.02.015.

[60] A. Luque, "Will we exceed 50% efficiency in photovoltaics?," *J. Appl. Phys.*, vol. 110, no. 3, Aug. 2011, doi: 10.1063/1.3600702.

[61] A. De Vos, "Detailed balance limit of the efficiency of tandem solar cells," *J. Phys. D. Appl. Phys.*, vol. 13, no. 5, pp. 839–846, May 1980, doi: 10.1088/0022-3727/13/5/018.

[62] G. Conibeer, "Third-generation photovoltaics," *Mater. Today*, vol. 10, no. 11, pp. 42–50, Nov. 2007, doi: 10.1016/S1369-7021(07)70278-X.

[63] X. Li, Y. Fang, and J. Zhao, "Optimizing Inorganic Cs4CuSb2Cl12/Cs2TiI6 Dual-Absorber Solar Cells: SCAPS-1D Simulations and Machine Learning," *Nanomaterials*, vol. 15, no. 16, 2025, doi: 10.3390/nano15161245.

[64] T. B. Dev, A. Srivani, S. Rajpoot, and S. Dhar, "Novel lead-free dual absorber based perovskite solar cells: efficient numerical harnessing towards high efficiency through SCAPS-1D," *J. Phys. Chem. Solids*, vol. 205, p. 112776, Oct. 2025, doi: 10.1016/j.jpcs.2025.112776.

[65] F. Ahmad, A. Lakhtakia, and P. B. Monk, "Double-absorber thin-film solar cell with 34% efficiency," *Appl. Phys. Lett.*, vol. 117, no. 3, 2020, doi: 10.1063/5.0017916.

[66] S. A. Moghadam Ziabari, A. Abdolahzadeh Ziabari, and S. J. Mousavi, "Efficiency enhancement of thin-film solar cell by implementation of double-absorber and BSF layers: the effect of thickness and carrier concentration," *J. Comput. Electron.*, vol. 21, no. 3, pp. 675–683, 2022, doi: 10.1007/s10825-022-01878-w.

[67] A. El Khalfi *et al.*, "Dual-Absorber Solar Cell Design and Simulation Based on Sb2Se3 and CZTGSe for High-Efficiency Solar Cells," *Langmuir*, vol. 40, no. 39, pp. 20352–20367, 2024, doi: 10.1021/acs.langmuir.4c01472.

[68] A. Kumar, M. Sujith, K. Valarmathi, R. Kumar, B. A. Al-Asbahi, and A. A. Ahmed, "Double-Absorber CZTS/Sb 2 Se 3 Architecture for High-Efficiency Solar-Cell Devices," *Phys. status solidi*, vol. 220, no. 11, Jun. 2023, doi: 10.1002/pssa.202200902.

[69] Mamta, K. K. Maurya, and V. N. Singh, "Sb2Se3/CZTS dual absorber layer based solar cell with 36.32 % efficiency: A numerical simulation," *J. Sci. Adv. Mater. Devices*, vol. 7, no. 2, p. 100445, 2022, doi: 10.1016/j.jsamd.2022.100445.

[70] S. H. Cheragee and M. J. Alam, "Device modelling and numerical analysis of high-efficiency double absorbers solar cells with diverse transport layer materials," *Results Opt.*, vol. 15, no. October 2023, p. 100647, 2024, doi: 10.1016/j.rio.2024.100647.

[71] T. AlZoubi, A. Moghrabi, M. Moustafa, and S. Yasin, "Efficiency boost of CZTS solar cells based on double-absorber architecture: Device modeling and analysis," *Sol. Energy*, vol. 225, no. July, pp. 44–52, 2021, doi: 10.1016/j.solener.2021.07.012.

[72] M. S. Rahman, S. Islam, A. Khandaker, T. Hossain, and M. J. Rashid, "Bilayer CZTS/Si absorber for obtaining highly efficient CZTS solar cell," *Sol. Energy*, vol. 230, no. November, pp. 1189–1198, 2021, doi: 10.1016/j.solener.2021.11.021.

[73] S. Bhattarai *et al.*, "Performance Improvement of Hybrid-Perovskite Solar Cells with Double Active Layer Design Using Extensive Simulation," *Energy and Fuels*, vol. 37, no. 21, pp. 16893–16903, 2023, doi: 10.1021/acs.energyfuels.3c02478.

[74] O. Ahmad *et al.*, "Modelling and numerical simulations of eco-friendly double absorber solar cell 'Spiro-OmeTAD/CIGS/MASnI3/CdS/ZnO' and its PV-module," *Org. Electron.*, vol. 117, no. December 2022, 2023, doi: 10.1016/j.orgel.2023.106781.

[75] T. Yadav, S. Yadav, and A. Sahu, "Comparative performance analysis of perovskite/CIGS-based double absorber layer solar cell with BaSi2 as a BSF layer," *J. Opt.*, vol. 53, no. 4, pp. 2922–2929, 2024, doi: 10.1007/s12596-023-01425-1.

[76] N. Selmane, A. Cheknane, F. Khemloul, M. H. S. Helal, and H. S. Hilal, "Cost-saving and performance-enhancement of CuInGaSe solar cells by adding CuZnSnSe as a second absorber," *Sol. Energy*, vol. 234, no. February, pp. 64–80, 2022, doi: 10.1016/j.solener.2022.01.072.

[77] M. Hasan Ali *et al.*, "Numerical analysis of FeSi2 based solar cell with PEDOT:PSS hole transport layer," *Mater. Today Commun.*, vol. 34, no. January, p. 105387, 2023, doi: 10.1016/j.mtcomm.2023.105387.

[78] Y. Makita, T. Ootsuka, Y. Fukuzawa, N. Otogawa, and H. Abe, "β -FeSi 2 as a Kankyo ( Environmentally Friendly ) semiconductor for solar cells in the space application," vol. 6197, pp. 1–14, 2006, doi: 10.1117/12.664009.

[79] A. S. W. Wong, G. W. Ho, and D. Z. Chi, "Understanding the Growth of β-FeSi[sub 2] Films for Photovoltaic Applications: A Study Using Transmission Electron Microscopy," *J. Electrochem. Soc.*, vol. 157, no. 8, p. H847, 2010, doi: 10.1149/1.3454741.

[80] M. Shaban, A. M. Bayoumi, D. Farouk, M. B. Saleh, and T. Yoshitake, "Evaluation of photovoltaic properties of nanocrystalline-FeSi 2 /Si heterojunctions," *Solid. State. Electron.*, vol. 123, pp. 111–118, Sep. 2016, doi: 10.1016/j.sse.2016.05.006.

[81] J. C. Z. Medina *et al.*, "Theoretical improvement of the energy conversion efficiency of an AZO/CdTe heterojunction solar cell to over 27% by incorporating FeSi2 as a second absorber layer," *Phys. Scr.*, vol. 99, no. 11, 2024, doi: 10.1088/1402-4896/ad86fb.

[82] M. F. Rahman *et al.*, "Design and numerical investigation of cadmium telluride (CdTe) and iron silicide (FeSi2) based double absorber solar cells to enhance power conversion efficiency," *AIP Adv.*, vol. 12, no. 10, Oct. 2022, doi: 10.1063/5.0108459.

[83] S. L. Liew *et al.*, "Thin Film Polycrystalline β-FeSi2/Si Heterojunction Solar Cells via Al Incorporation and Rapid Thermal Processing," *Energy Procedia*, vol. 15, pp. 305–311, 2012, doi: 10.1016/j.egypro.2012.02.036.

[84] A. Kumar *et al.*, "Integration of β-FeSi2 with poly-Si on glass for thin-film photovoltaic applications," *RSC Adv.*, vol. 3, no. 21, p. 7733, 2013, doi: 10.1039/c3ra41156g.

[85] M. M. A. Moon, M. H. Ali, M. F. Rahman, J. Hossain, and A. B. M. Ismail, "Design and Simulation of FeSi2-Based Novel Heterojunction Solar Cells for Harnessing Visible and Near-Infrared Light," *Phys. Status Solidi Appl. Mater. Sci.*, vol. 217, no. 6, pp. 1–12, 2020, doi: 10.1002/pssa.201900921.

[86] F. Alharbi, J. D. Bass, A. Salhi, A. Alyamani, H.-C. Kim, and R. D. Miller, "Abundant non-toxic materials for thin film solar cells: Alternative to conventional materials," *Renew. Energy*, vol. 36, no. 10, pp. 2753–2758, Oct. 2011, doi: 10.1016/j.renene.2011.03.010.

[87] J. Yuan, H. Shen, L. Lu, H. Huang, and X. He, "Effects of emitter parameters and recombination mechanisms on the performance of β-FeSi2/c-Si heterojunction solar cells," *Phys. B Condens. Matter*, vol. 405, no. 21, pp. 4565–4569, Nov. 2010, doi: 10.1016/j.physb.2010.08.039.

[88] G. K. Dalapati *et al.*, "cells and the effects of interfacial engineering," vol. 2, no. 2011, pp. 2012–2015, 2014, doi: 10.1063/1.3536523.

[89] Y. Gao, H. W. Liu, Y. Lin, and G. Shao, "Computational design of high efficiency FeSi2 thin-film solar cells," *Thin Solid Films*, vol. 519, no. 24, pp. 8490–8495, Oct. 2011, doi: 10.1016/j.tsf.2011.05.030.

[90] B. Sultana *et al.*, "A novel design and optimization of Si based high performance double absorber heterojunction solar cell," *Mater. Sci. Eng. B*, vol. 304, no. April, p. 117360, 2024, doi: 10.1016/j.mseb.2024.117360.

[91] P. Parathraju and P. Umasankar, "Performance evaluation of ultrathin CdTe-based solar cells with dual absorbers via SCAPS-1D simulation," *Sci. Rep.*, vol. 15, no. 1, pp. 1–20, 2025, doi: 10.1038/s41598-025-12006-6.

[92] M. Burgelman, P. Nollet, and S. Degrave, "Modelling polycrystalline semiconductor solar cells," *Thin Solid Films*, vol. 361–362, pp. 527–532, Feb. 2000, doi: 10.1016/S0040-6090(99)00825-1.

[93] M. Burgelman, J. Verschraegen, S. Degrave, and P. Nollet, "Modeling thin-film PV devices," *Prog. Photovoltaics Res. Appl.*, vol. 12, no. 2–3, pp. 143–153, Mar. 2004, doi: 10.1002/pip.524.

[94] M. Burgelman, "SCAPS Manual." Accessed: Nov. 05, 2025. [Online]. Available: https://scaps.elis.ugent.be/SCAPS manual most recent.pdf

[95] H. Bencherif and M. Khalid Hossain, "Design and numerical investigation of efficient (FAPbI3)1−x(CsSnI3)x perovskite solar cell with optimized performances," *Sol. Energy*, vol. 248, pp. 137–148, Dec. 2022, doi: 10.1016/j.solener.2022.11.012.

[96] H. Bencherif *et al.*, "Performance enhancement of (FAPbI3)1-x(MAPbBr3)x perovskite solar cell with an

optimized design," *Micro and Nanostructures*, vol. 171, p. 207403, Nov. 2022, doi: 10.1016/j.micrna.2022.207403.

[97] M. F. Rahman *et al.*, "Concurrent investigation of antimony chalcogenide (Sb2Se3 and Sb2S3)-based solar cells with a potential WS2 electron transport layer," *Heliyon*, vol. 8, no. 12, p. e12034, Dec. 2022, doi: 10.1016/j.heliyon.2022.e12034.

[98] M. K. Hossain *et al.*, "Numerical simulation and optimization of a CsPbI 3 -based perovskite solar cell to enhance the power conversion efficiency," *New J. Chem.*, vol. 47, no. 10, pp. 4801–4817, 2023, doi: 10.1039/D2NJ06206B.

[99] M. K. Hossain *et al.*, "Numerical Analysis in DFT and SCAPS-1D on the Influence of Different Charge Transport Layers of CsPbBr 3 Perovskite Solar Cells," *Energy & Fuels*, vol. 37, no. 8, pp. 6078–6098, Apr. 2023, doi: 10.1021/acs.energyfuels.3c00035.

[100] J. Li *et al.*, "Unveiling microscopic carrier loss mechanisms in 12% efficient Cu2ZnSnSe4 solar cells," *Nat. Energy*, vol. 7, no. 8, pp. 754–764, Jul. 2022, doi: 10.1038/s41560-022-01078-7.

[101] S. M. Seyed-Talebi, M. Mahmoudi, and C.-H. Lee, "A Comprehensive Study of CsSnI3-Based Perovskite Solar Cells with Different Hole Transporting Layers and Back Contacts," *Micromachines*, vol. 14, no. 8, p. 1562, Aug. 2023, doi: 10.3390/mi14081562.

[102] Marc Burgelman, "Models for the optical absorption of materials in SCAPS." [Online]. Available: https://scaps.elis.ugent.be/SCAPS Application Note Absorption Models.pdf

[103] C.B.Honsberg and S.G.Bowden, "Photovoltaics Education Website." [Online]. Available: www.pveducation.org

[104] J. Patel, R. K. Sharme, M. A. Quijada, and M. M. Rana, "A Review of Transparent Conducting Films (TCFs): Prospective ITO and AZO Deposition Methods and Applications," *Nanomaterials*, vol. 14, no. 24, 2024, doi: 10.3390/nano14242013.

[105] H. Heffner, M. Soldera, and A. F. Lasagni, "Optoelectronic performance of indium tin oxide thin films structured by sub-picosecond direct laser interference patterning," *Sci. Rep.*, vol. 13, no. 1, pp. 1–12, 2023, doi: 10.1038/s41598-023-37042-y.

[106] C. Chang, S. Panigrahy, and D. K. Das, "Performance analysis of ultra-thin CIGS solar cells with ZnS/CdS/ZnSe buffer layers," *J. Opt.*, 2024, doi: 10.1007/s12596-024-01818-w.

[107] Y. Zeng *et al.*, "Comparative Study of TiO 2 and CdS as the Electron Transport Layer for Sb 2 S 3 Solar Cells," *Sol. RRL*, vol. 6, no. 10, Oct. 2022, doi: 10.1002/solr.202200435.

[108] K. S. Cho *et al.*, "Optimal CdS buffer thickness to form high-quality CdS/Cu(In,Ga)Se2 junctions in solar cells without plasma damage and shunt paths," *ACS Omega*, vol. 5, no. 37, pp. 23983–23988, 2020, doi: 10.1021/acsomega.0c03268.

[109] C. Zhao *et al.*, "Advances in CIGS thin film solar cells with emphasis on the alkali element post-deposition treatment," *Mater. Reports Energy*, vol. 3, no. 3, p. 100214, 2023, doi: 10.1016/j.matre.2023.100214.

[110] K. Zeghdar *et al.*, "Optimizing photovoltaic efficiency in CZTS solar cells by investigating the role of different advanced materials as back surface field layer," *Sci. Rep.*, vol. 15, no. 1, p. 25294, Jul. 2025, doi: 10.1038/s41598-025-10958-3.

[111] M. R. Sultana, B. Islam, and S. R. Al Ahmed, "Modeling and Performance Analysis of Highly Efficient Copper Indium Gallium Selenide Solar Cell with Cu 2 O Hole Transport Layer Using Solar Cell Capacitance Simulator in One Dimension," *Phys. status solidi*, vol. 219, no. 5, Mar. 2022, doi: 10.1002/pssa.202100512.

[112] S. Chatterjee and A. J. Pal, "Introducing Cu 2 O Thin Films as a Hole-Transport Layer in Efficient Planar Perovskite Solar Cell Structures," *J. Phys. Chem. C*, vol. 120, no. 3, pp. 1428–1437, Jan. 2016, doi: 10.1021/acs.jpcc.5b11540.

[113] W. Xie *et al.*, "Numerical investigation on the performance of heterojunction solar cells with Cu2O as the hole transport layer and Cu2MoSnS4 as the absorption layer," *Phys. Lett. A*, vol. 528, p. 130029, Dec. 2024, doi: 10.1016/j.physleta.2024.130029.

[114] R. Ranjan *et al.*, "SCAPS study on the effect of various hole transport layer on highly efficient 31.86% eco-friendly CZTS based solar cell," *Sci. Rep.*, vol. 13, no. 1, pp. 1–16, 2023, doi: 10.1038/s41598-023-44845-6.

[115] M. Stössel, J. Staudigel, F. Steuber, J. Simmerer, and A. Winnacker, "Impact of the cathode metal work function on the performance of vacuum-deposited organic light emitting-devices," *Appl. Phys. A Mater. Sci. Process.*, vol. 68, no. 4, pp. 387–390, 1999, doi: 10.1007/s003390050910.

[116] S. Karthick, S. Velumani, and J. Bouclé, "Chalcogenide BaZrS3 perovskite solar cells: A numerical simulation and analysis using SCAPS-1D," *Opt. Mater. (Amst).*, vol. 126, no. March, pp. 1–10, 2022, doi: 10.1016/j.optmat.2022.112250.

[117] M. El-Mrabet et al., "Advancing solar cell efficiency: insights from cesium lead halide perovskite analysis," *J. Mater. Sci. Mater. Electron.*, vol. 36, no. 27, p. 1759, Sep. 2025, doi: 10.1007/s10854-025-15827-3.

[118] P. Sawicka-Chudy, M. Sibiński, E. Rybak-Wilusz, M. Cholewa, G. Wisz, and R. Yavorskyi, "Review of the development of copper oxides with titanium dioxide thin-film solar cells," *AIP Adv.*, vol. 10, no. 1, Jan. 2020, doi: 10.1063/1.5125433.

[119] S. R. I. Biplab, M. H. Ali, M. M. A. Moon, M. F. Pervez, M. F. Rahman, and J. Hossain, "Performance enhancement of CIGS-based solar cells by incorporating an ultrathin BaSi2 BSF layer," *J. Comput. Electron.*, vol. 19, no. 1, pp. 342–352, Mar. 2020, doi: 10.1007/s10825-019-01433-0.

[120] M. M. A. Moon, M. H. Ali, M. F. Rahman, A. Kuddus, J. Hossain, and A. B. M. Ismail, "Investigation of thin-film p -BaSi 2 / n -CdS heterostructure towards semiconducting silicide based high efficiency solar cell," *Phys. Scr.*, vol. 95, no. 3, p. 035506, Mar. 2020, doi: 10.1088/1402-4896/ab49e8.

[121] M. D. Haque, M. H. Ali, and A. Z. M. T. Islam, "Efficiency enhancement of WSe2 heterojunction solar cell with CuSCN as a hole transport layer: A numerical simulation approach," *Sol. Energy*, vol. 230, pp. 528–537, Dec. 2021, doi: 10.1016/j.solener.2021.10.054.

[122] P. Hartnagel and T. Kirchartz, "Understanding the Light-Intensity Dependence of the Short-Circuit Current of Organic Solar Cells," vol. 2000116, pp. 1–11, 2020, doi: 10.1002/adts.202000116.

[123] S. H. Cheragee and M. J. Alam, "Device modeling and numerical study of a double absorber solar cell using a variety of electron transport materials," *Heliyon*, vol. 9, no. 7, p. e18265, 2023, doi: 10.1016/j.heliyon.2023.e18265.

[124] T. Ameri, G. Dennler, C. Lungenschmied, and C. J. Brabec, "Organic tandem solar cells: A review," *Energy Environ. Sci.*, vol. 2, no. 4, p. 347, 2009, doi: 10.1039/b817952b.

[125] A. Mohandes, A. C. Roy, N. Rahman, M. Amami, S. Ezzine, and M. Ferdous Rahman, *A numerical strategy to achieving efficiency exceeding 27% with a novel dual absorber perovskite solar cell using BaZrSe3 and CsPbI3*, vol. 57, no. 1. Springer US, 2025. doi: 10.1007/s11082-025-08043-0.

[126] M. H. Tonmoy, S. N. Shiddique, A. T. Abir, and J. Hossain, "Design and optimization of a high efficiency CdTe–FeSi2 based double-junction two-terminal tandem solar cell," *Heliyon*, vol. 10, no. 6, p. e27994, 2024, doi: 10.1016/j.heliyon.2024.e27994.

[127] M. M. Khatun, A. Hosen, and S. R. Al Ahmed, "Evaluating the performance of efficient Cu2NiSnS4 solar cell—A two stage theoretical attempt and comparison to experiments," *Heliyon*, vol. 9, no. 10, p. e20603, Oct. 2023, doi: 10.1016/j.heliyon.2023.e20603.

[128] M. D. Wanda, S. Ouédraogo, F. Tchoffo, F. Zougmoré, and J. M. B. Ndjaka, "Numerical Investigations and Analysis of Cu2ZnSnS4 Based Solar Cells by SCAPS-1D," *Int. J. Photoenergy*, vol. 2016, 2016, doi: 10.1155/2016/2152018.

[129] A. Hosen, M. S. Mian, and S. R. Al Ahmed, "Simulating the performance of a highly efficient CuBi2O4-based thin-film solar cell," *SN Appl. Sci.*, vol. 3, no. 5, pp. 1–13, 2021, doi: 10.1007/s42452-021-04554-z.

[130] R. K. Shukla et al., "Simulation study of solar cell with a double absorber layers of perovskites material using lead and lead-free material," *J. Opt.*, vol. 53, no. 5, pp. 4477–4486, 2024, doi: 10.1007/s12596-024-01678-4.

[131] R. K. Shukla, A. Srivastava, N. Wadhwani, S. Shukla, and N. Singh, "Modeling and simulation of lead-free, formamidinium germanium-antimony halide (FA4GeSbCl12) based solar cell," *J. Opt.*, 2025, doi: 10.1007/s12596-025-02599-6.

[132] A. Tara, V. Bharti, S. Sharma, and R. Gupta, "Device simulation of FASnI3 based perovskite solar cell with Zn(O0.3, S0.7) as electron transport layer using SCAPS-1D," *Opt. Mater. (Amst).*, vol. 119, no. May, p. 111362, 2021, doi: 10.1016/j.optmat.2021.111362.

[133] M. C. Islam, B. K. Mondal, T. Ahmed, M. A. H. Pappu, S. K. Mostaque, and J. Hossain, "Design of a highly efficient n-CdS/p-AgGaTe 2 /p+-SnS double-heterojunction thin film solar cell," *Eng. Res. Express*, vol. 5, no. 2, p. 025056, Jun. 2023, doi: 10.1088/2631-8695/acd98a.

[134] A. Ali et al., "Design and numerical simulation of lead-free inorganic perovskites (Cs2BiAgI6) solar cell through computation method," *Heliyon*, vol. 10, no. 14, p. e33922, 2024, doi: 10.1016/j.heliyon.2024.e33922.

[135] L. Wang et al., "Defects in kesterite materials towards high-efficiency solar cells: origin, impact,

characterization, and engineering," *J. Mater. Chem. A*, vol. 12, no. 38, pp. 25643–25677, 2024, doi: 10.1039/D4TA03883E.

[136] R. Fonoll-Rubio *et al.*, "Insights into interface and bulk defects in a high efficiency kesterite-based device," *Energy Environ. Sci.*, vol. 14, no. 1, pp. 507–523, 2021, doi: 10.1039/D0EE02004D.

[137] M. S. Jamal *et al.*, "Effect of defect density and energy level mismatch on the performance of perovskite solar cells by numerical simulation," *Optik (Stuttg).*, vol. 182, pp. 1204–1210, Apr. 2019, doi: 10.1016/j.ijleo.2018.12.163.

[138] G. Vishnupriya and P. Sathya, "Computational study of a novel combination of dual-absorber structured perovskite solar cell with theoretical efficiency of 36.37%," *Front. Energy Res.*, vol. 13, no. July, pp. 1–12, 2025, doi: 10.3389/fenrg.2025.1631201.

[139] S. Islam *et al.*, " Next-generation dual absorber solar cell design with Ca 3 AsI 3 and Sr 3 PBr 3 perovskites and MoO 3 HTL achieves superior efficiency above 29% ," *Energy Adv.*, 2025, doi: 10.1039/d5ya00137d.

[140] M. M. Haque *et al.*, "Study on the interface defects of eco-friendly perovskite solar cells," *Sol. Energy*, vol. 247, pp. 96–108, Nov. 2022, doi: 10.1016/j.solener.2022.10.024.

[141] F. Rahman, A. Rahman, and M. Z. Bani-fwaz, "RSC Advances Advancing photovoltaics with Cs 2 NaInI 6 -based," *RSC Adv.*, vol. 15, pp. 38122–38133, 2025, doi: 10.1039/D5RA05885F.

[142] F. T. Zahra, M. M. Hasan, M. B. Hossen, and M. R. Islam, "Deep insights into the optoelectronic properties of AgCdF3-based perovskite solar cell using the combination of DFT and SCAPS-1D simulation," *Heliyon*, vol. 10, no. 13, p. e33096, 2024, doi: 10.1016/j.heliyon.2024.e33096.

[143] R. Tathe, K. B. Chavan, S. Chaure, and N. B. Chaure, "Studies on the effect of metal contact work function on performance of CZTSSe thin film solar cell: A numerical analysis using SCAPS 1D," *Next Res.*, vol. 2, no. 2, p. 100265, 2025, doi: 10.1016/j.nexres.2025.100265.

[144] C. Aliani, M. Krichen, and A. Zouari, "Effect of the front-metal work function on the performance of a-Si:H(n+)/a-Si:H(i)/c-Si(p) heterojunction solar cells," *J. Comput. Electron.*, vol. 18, no. 2, pp. 576–583, 2019, doi: 10.1007/s10825-019-01324-4.

[145] J. Madan, S. Garg, K. Gupta, S. Rana, A. Manocha, and R. Pandey, "Numerical simulation of charge transport layer free perovskite solar cell using metal work function shifted contacts," *Optik (Stuttg).*, vol. 202, no. August 2019, p. 163646, 2020, doi: 10.1016/j.ijleo.2019.163646.

[146] V. T. Babu *et al.*, "Role of front and back contacts in the performance of TiO2/CuO heterojunction solar cells," *Mater. Today Proc.*, vol. 65, pp. 408–415, 2022, doi: 10.1016/j.matpr.2022.07.199.

[147] P. K. Dakua *et al.*, "Optimization of efficiency of CZTS-based solar cell through exclusive BSF layer engineering method," *J. Phys. Chem. Solids*, vol. 193, no. March, p. 112156, 2024, doi: 10.1016/j.jpcs.2024.112156.

[148] E. N. Vincent Mercy, D. Srinivasan, and L. Marasamy, "Emerging BaZrS 3 and Ba(Zr,Ti)S 3 Chalcogenide Perovskite Solar Cells: A Numerical Approach Toward Device Engineering and Unlocking Efficiency," *ACS Omega*, vol. 9, no. 4, pp. 4359–4376, Jan. 2024, doi: 10.1021/acsomega.3c06627.

[149] C. Lin *et al.*, "The Investigation of the Influence of a Cu2O Buffer Layer on Hole Transport Layers in MAPbI3-Based Perovskite Solar Cells," *Materials (Basel).*, vol. 15, no. 22, 2022, doi: 10.3390/ma15228142.

[150] A. Kumar and A. D. Thakur, "Role of contact work function, back surface field, and conduction band offset in Cu2ZnSnS4 solar cell," *Jpn. J. Appl. Phys.*, vol. 57, no. 8, 2018, doi: 10.7567/JJAP.57.08RC05.

[151] M. Samiul Islam *et al.*, "Defect Study and Modelling of SnX3-Based Perovskite Solar Cells with SCAPS-1D," *Nanomaterials*, vol. 11, no. 5, p. 1218, May 2021, doi: 10.3390/nano11051218.

[152] M. Becker and M. Wark, "Sequentially Deposited Compact and Pinhole-Free Perovskite Layers via Adjusting the Permittivity of the Conversion Solution," *Zeitschrift für Naturforsch. A*, vol. 74, no. 8, pp. 655–663, Aug. 2019, doi: 10.1515/zna-2019-0141.

[153] R. Singh, S. Sandhu, and J.-J. Lee, "Elucidating the effect of shunt losses on the performance of mesoporous perovskite solar cells," *Sol. Energy*, vol. 193, pp. 956–961, Nov. 2019, doi: 10.1016/j.solener.2019.10.018.

[154] D. Saranin *et al.*, "Copper Iodide Interlayer for Improved Charge Extraction and Stability of Inverted Perovskite Solar Cells," *Materials (Basel).*, vol. 12, no. 9, p. 1406, Apr. 2019, doi: 10.3390/ma12091406.

[155] Z. Bi *et al.*, "High Shunt Resistance SnO 2 -PbO Electron Transport Layer for Perovskite Solar Cells Used in Low Lighting Applications," *Adv. Sustain. Syst.*, vol. 5, no. 11, Nov. 2021, doi: 10.1002/adsu.202100120.

[156] T. I. Taseen, M. Julkarnain, and A. Z. M. T. Islam, "Design and simulation of nitrogenated holey graphene (C2N) based heterostructure solar cell by SCAPS-1D," *Heliyon*, vol. 10, no. 1, p. e23197, Jan. 2024, doi: 10.1016/j.heliyon.2023.e23197.

[157] Y. P. Varshni, "Temperature dependence of the energy gap in semiconductors," *Physica*, vol. 34, no. 1, pp. 149–154, Jan. 1967, doi: 10.1016/0031-8914(67)90062-6.

[158] S. R. Al Ahmed, A. Sunny, and S. Rahman, "Performance enhancement of Sb2Se3 solar cell using a back surface field layer: A numerical simulation approach," *Sol. Energy Mater. Sol. Cells*, vol. 221, p. 110919, Mar. 2021, doi: 10.1016/j.solmat.2020.110919.

[159] P. Singh and N. M. Ravindra, "Temperature dependence of solar cell performance - An analysis," *Sol. Energy Mater. Sol. Cells*, vol. 101, no. October, pp. 36–45, 2012, doi: 10.1016/j.solmat.2012.02.019.

[160] A. Ghosh *et al.*, "Numerical analysis and device modelling of a lead-free Sr3PI3/Sr3SbI3 double absorber solar cell for enhanced efficiency," *RSC Adv.*, vol. 14, no. 36, pp. 26437–26456, 2024, doi: 10.1039/d4ra05079g.

[161] J. H. Kim, K. J. Moon, J. M. Kim, D. Lee, and S. H. Kim, "Effects of various light-intensity and temperature environments on the photovoltaic performance of dye-sensitized solar cells," *Sol. Energy*, vol. 113, pp. 251–257, Mar. 2015, doi: 10.1016/j.solener.2015.01.012.

[162] M. T. Bin Kashem and S. A. Esha, "Advancing lead-free all-inorganic NaSnCl3 based perovskite solar cell for high efficiency: Computational optimization of charge transport layers and material parameters," *Mater. Today Commun.*, vol. 47, p. 113192, Jul. 2025, doi: 10.1016/j.mtcomm.2025.113192.

[163] L. Chen *et al.*, "Synthesis and characterization of earth-abundant Cu 2 MnSnS 4 thin films using a non-toxic solution-based technique," *RSC Adv.*, vol. 5, no. 102, pp. 84295–84302, 2015, doi: 10.1039/C5RA14595C.

[164] W. Wang *et al.*, "A general oxide-based preparation strategy for Cu2MSnS4 (M: Zn, Mn, Cd) thin films and relevant solar cells," *Mater. Lett.*, vol. 214, pp. 170–173, 2018, doi: 10.1016/j.matlet.2017.12.001.

[165] T. A. Chowdhury, "Numerical analysis to enhance efficiency of Cu 2 MgSnS 4 -based solar cell by inserting V 2 O 5 back surface field," *Phys. Scr.*, vol. 100, no. 4, p. 045951, Apr. 2025, doi: 10.1088/1402-4896/adbdfa.

[166] Z. Liu, S. Wang, and N. Otogawa, "A thin-film solar cell of high-quality b -FeSi 2 / Si heterojunction prepared by sputtering," vol. 90, pp. 276–282, 2006, doi: 10.1016/j.solmat.2005.03.014.

[167] C. Wen and Y. Lin, "Journal of Physics and Chemistry of Solids Cu 2 O hole transport layer optimization for Sb 2 ( S , Se ) 3 solar cells : combining machine learning prediction with experimental validation," vol. 209, no. September 2025, pp. 1–11, 2026, doi: 10.1016/j.jpcs.2025.113298.

[168] A. S. Hassanien and I. M. El Radaf, "Optical characterizations of quaternary Cu2MnSnS4 thin films: Novel synthesis process of film samples by spray pyrolysis technique," *Phys. B Condens. Matter*, vol. 585, p. 412110, May 2020, doi: 10.1016/j.physb.2020.412110.

[169] H. Katsumata, Y. Makita, N. Kobayashi, M. Hasegawa, H. Shibata, and S. Uekusa, "Synthesis of β-FeSi2 for optical applications by Fe triple-energy ion implantation into Si(100) and Si(111) substrates," *Thin Solid Films*, vol. 281–282, pp. 252–255, Aug. 1996, doi: 10.1016/0040-6090(96)08645-2.

[170] H. Katsumata *et al.*, "Optical absorption and photoluminescence studies of β-FeSi2 prepared by heavy implantation of Fe+ ions into Si," *J. Appl. Phys.*, vol. 80, no. 10, pp. 5955–5962, Nov. 1996, doi: 10.1063/1.363591.

[171] H. Kakemoto, Y. Makita, Y. Kino, S. Sakuragi, and T. Tsukamoto, "Small polaron of β-FeSi2 obtained from optical measurements," *Thin Solid Films*, vol. 381, no. 2, pp. 251–255, Jan. 2001, doi: 10.1016/S0040-6090(00)01752-1.

[172] Z. Liu, M. Watanabe, and M. Hanabusa, "Electrical and photovoltaic properties of iron-silicide/silicon heterostructures formed by pulsed laser deposition," *Thin Solid Films*, vol. 381, no. 2, pp. 262–266, Jan. 2001, doi: 10.1016/S0040-6090(00)01754-5.

[173] S. Wang *et al.*, "Formation of β-FeSi 2 Microstructures by Reactive Ion Etching Using SF 6 Gas," *Jpn. J. Appl. Phys.*, vol. 43, no. 8R, p. 5245, Aug. 2004, doi: 10.1143/JJAP.43.5245.

[174] J. Han *et al.*, "Solar Energy Materials & Solar Cells Optimized chemical bath deposited CdS layers for the improvement of CdTe solar cells," *Sol. Energy Mater. Sol. Cells*, vol. 95, no. 3, pp. 816–820, 2011, doi: 10.1016/j.solmat.2010.10.027.

[175] C. Schwartz *et al.*, "Electronic structure study of the CdS buffer layer in CIGS solar cells by X-ray absorption spectroscopy: Experiment and theory," *Sol. Energy Mater. Sol. Cells*, vol. 149, pp. 275–283, May 2016, doi: 10.1016/j.solmat.2016.01.043.

[176] J. Zhu, Z. Xu, G. Fan, S. Lee, Y. Li, and J. Tang, "Inverted polymer solar cells with atomic layer deposited CdS film as an electron collection layer," *Org. Electron.*, vol. 12, no. 12, pp. 2151–2158, 2011, doi: 10.1016/j.orgel.2011.09.007.

[177] J. R. Bakke, H. J. Jung, J. T. Tanskanen, R. Sinclair, and S. F. Bent, "Atomic Layer Deposition of CdS Films," no. 3, pp. 4669–4678, 2010, doi: 10.1021/cm100874f.

[178] A. Ashok, G. Regmi, A. Romero-Núñez, M. Solis-López, S. Velumani, and H. Castaneda, "Comparative studies of CdS thin films by chemical bath deposition techniques as a buffer layer for solar cell applications," *J. Mater. Sci. Mater. Electron.*, vol. 31, no. 10, pp. 7499–7518, May 2020, doi: 10.1007/s10854-020-03024-3.

[179] Y. Hashimoto, N. Kohara, T. Negami, N. Nishitani, and T. Wada, "Chemical bath deposition of Cds buffer layer for GIGS solar cells," *Sol. Energy Mater. Sol. Cells*, vol. 50, no. 1–4, pp. 71–77, Jan. 1998, doi: 10.1016/S0927-0248(97)00124-4.

[180] R. P. Wijesundera, "Fabrication of the CuO / Cu2O heterojunction using an electrodeposition technique for solar cell applications Fabrication of the CuO / Cu2O heterojunction using an electrodeposition technique for solar cell applications," no. March 2010, 2014, doi: 10.1088/0268-1242/25/4/045015.

[181] M. H. Tran, J. Y. Cho, S. Sinha, M. G. Gang, and J. Heo, "NU SC," *Thin Solid Films*, p. #pagerange#, 2018, doi: 10.1016/j.tsf.2018.07.023.

[182] S. Farid, U. Farhad, D. Cherns, J. A. Smith, N. A. Fox, and D. J. Fermín, "Pulsed laser deposition of single phase n- and p-type Cu 2 O thin fi lms with low resistivity," *Mater. Des.*, vol. 193, p. 108848, 2020, doi: 10.1016/j.matdes.2020.108848.

[183] C. V. Kartha *et al.*, "Insights into Cu2O thin film absorber via pulsed laser deposition," *Ceram. Int.*, vol. 48, no. 11, pp. 15274–15281, Jun. 2022, doi: 10.1016/j.ceramint.2022.02.061.

[184] V. Pakštas *et al.*, "Improvement of CZTSSe film quality and superstrate solar cell performance through optimized post - deposition annealing," *Sci. Rep.*, pp. 1–9, 2022, doi: 10.1038/s41598-022-20670-1.

[185] S. Chamekh, N. Khemiri, and M. Kanzari, "Effect of annealing under different atmospheres of CZTS thin films as absorber layer for solar cell application," *SN Appl. Sci.*, vol. 2, no. 9, pp. 1–8, 2020, doi: 10.1007/s42452-020-03287-9.

[186] J. Xu *et al.*, "Low-Temperature Annealing of CdS:In/Cu 2 ZnSn(S,Se) 4 Heterojunction Boosting 14.5% Efficiency Kesterite Solar Cells," *ACS Energy Lett.*, vol. 9, no. 10, pp. 4939–4946, Oct. 2024, doi: 10.1021/acsenergylett.4c02118.

[187] M. A. Olgar, A. O. Sarp, A. Seyhan, and R. Zan, "Impact of stacking order and annealing temperature on properties of CZTS thin fi lms and solar cell performance," *Renew. Energy*, vol. 179, pp. 1865–1874, 2021, doi: 10.1016/j.renene.2021.08.023.

[188] F. Werner, B. Veith-wolf, M. Melchiorre, F. Babbe, J. Schmidt, and S. Siebentritt, "Ultra-thin passivation layers in Cu ( In , Ga ) Se 2 thin-film solar cells : full-area passivated front contacts and their impact on bulk doping," *Sci. Rep.*, pp. 1–14, 2020, doi: 10.1038/s41598-020-64448-9.

[189] P. Ferdowsi, E. O. Martinez, S. S. Alonso, and U. Steiner, "Ultrathin polymeric films for interfacial passivation in wide band - gap perovskite solar cells," *Sci. Rep.*, pp. 1–10, 2020, doi: 10.1038/s41598-020-79348-1.

[190] K. Su, W. Chen, Y. Huang, G. Yang, and K. G. Brooks, "In Situ Graded Passivation via Porphyrin Derivative with Enhanced Photovoltage and Fill Factor in Perovskite Solar Cells," vol. 2100964, pp. 1–8, 2022, doi: 10.1002/solr.202100964.

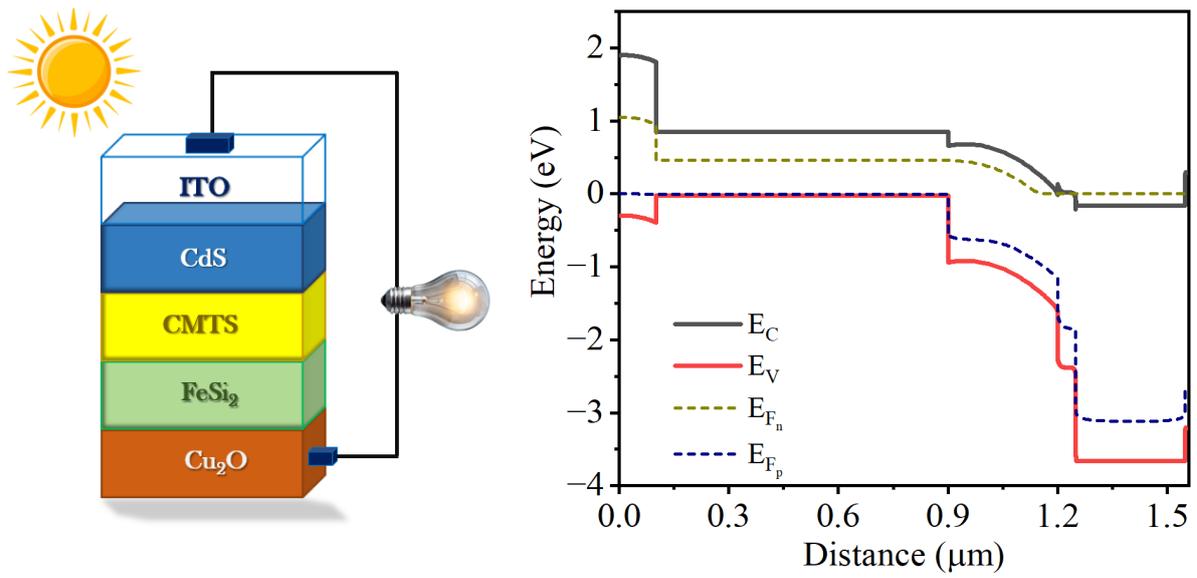

Fig. 1. (a) Schematic diagram of the dual-absorber solar cell configuration used in the simulations, and (b) corresponding energy band diagram of the structure.

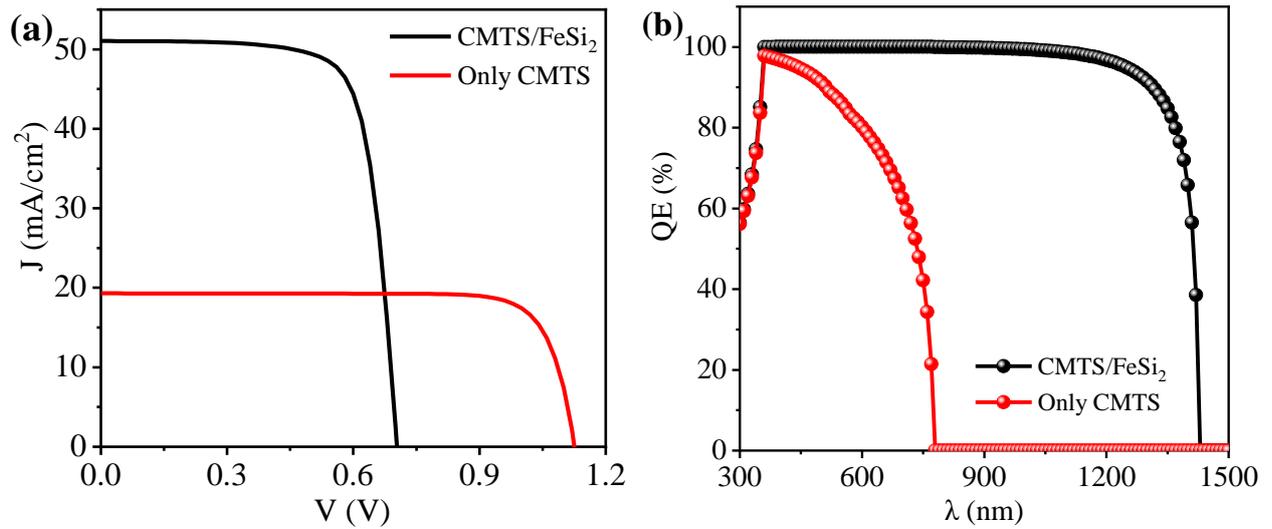

Fig. 2. (a) Current density (J) − voltage (V) characteristics and (b) quantum efficiency (QE) versus wavelength (λ) for the single-absorber (only CMTS) solar cell and the dual-absorber (CMTS/FeSi$_2$) solar cell. QE represents the ratio between the number of charge carriers collected by the solar cell and the number of incident photons of a specific energy [103].

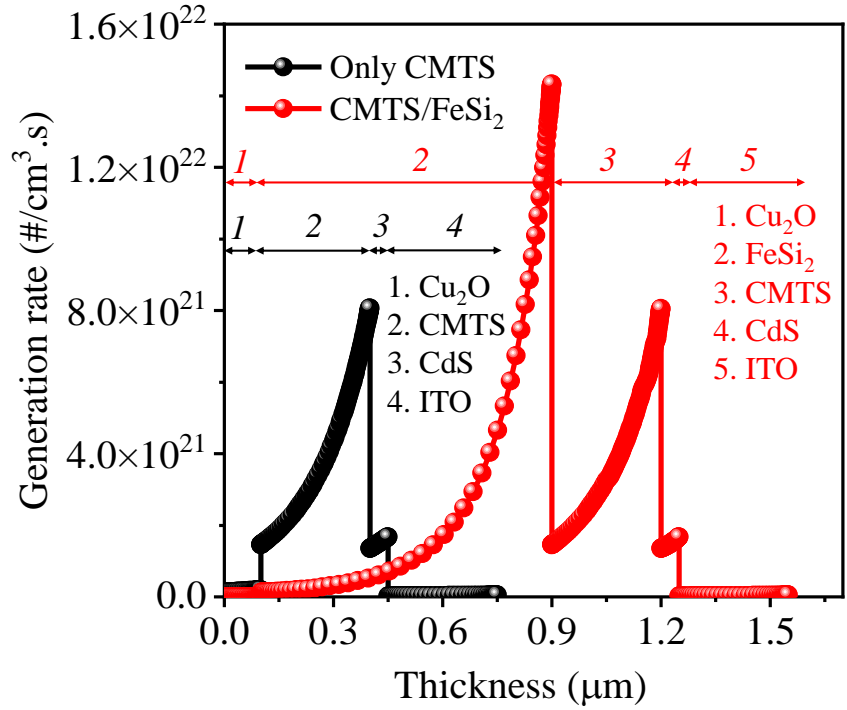

Fig. 3. Generation rate profile across different layers of the single-absorber (only CMTS) and dual-absorber (CMTS/FeSi$_2$) solar cell structures. The arrows indicate the thickness of each layer, with corresponding numbers shown above. Light is incident from the ITO side in both devices. The higher generation rates confirm that CMTS and FeSi$_2$ act as the absorber layers. The additional generation observed in the dual-absorber device leads to a higher J$_{SC}$, as J$_{SC}$ is directly proportional to the total photogeneration rate within the active layers [122].

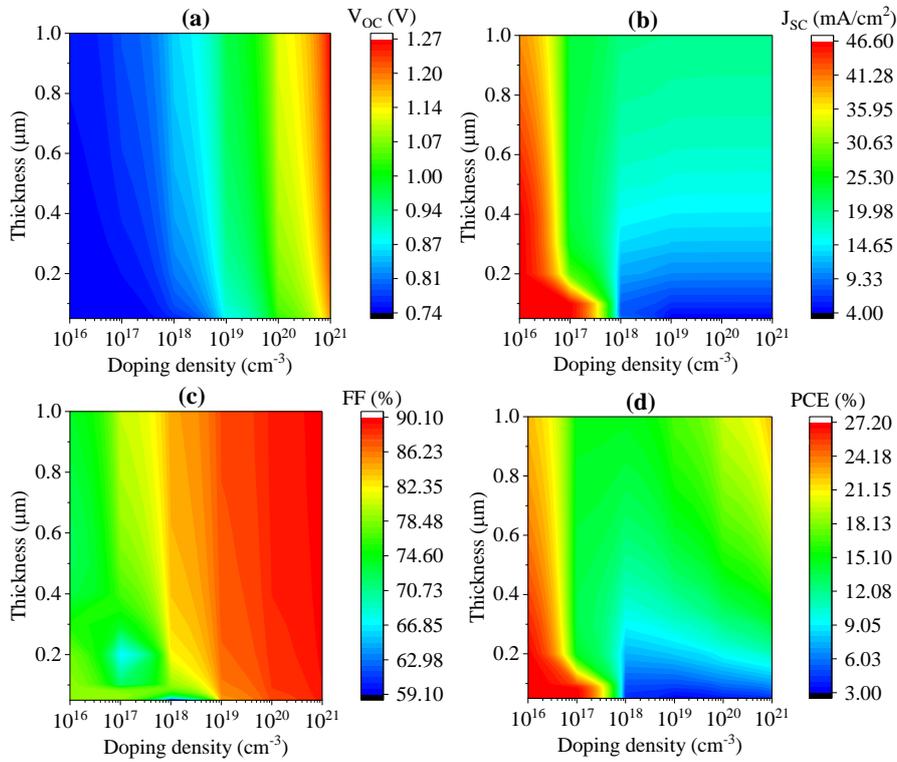

Fig. 4. Effect of thickness and doping density of the primary absorber, CMTS on V$_{OC}$, J$_{SC}$, FF, and PCE of the device. All other material parameters were kept fixed at their default values, as listed in Tables 1 and 2.

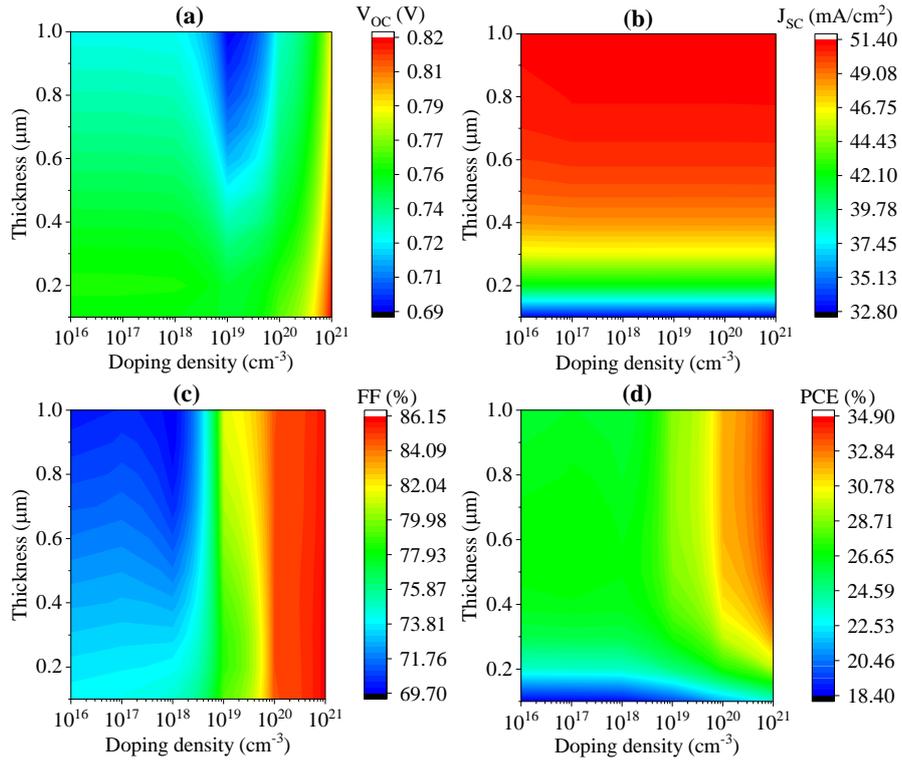

Fig. 5. Effect of thickness and doping density of the secondary absorber, FeSi$_2$ on V$_{OC}$, J$_{SC}$, FF, and PCE of the device. All other material parameters were kept fixed at their default values, as listed in Tables 1 and 2.

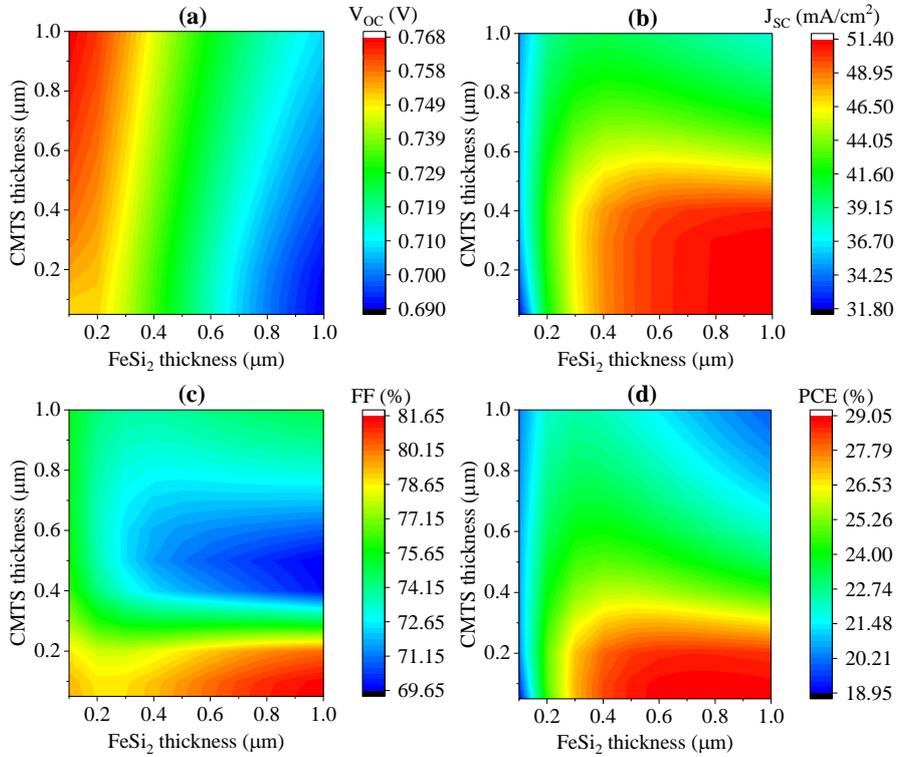

Fig. 6. Combined effect of CMTS and FeSi$_2$ thicknesses on V$_{OC}$, J$_{SC}$, FF, and PCE of the device. All other material parameters were kept fixed at their default values, as listed in Tables 1 and 2.

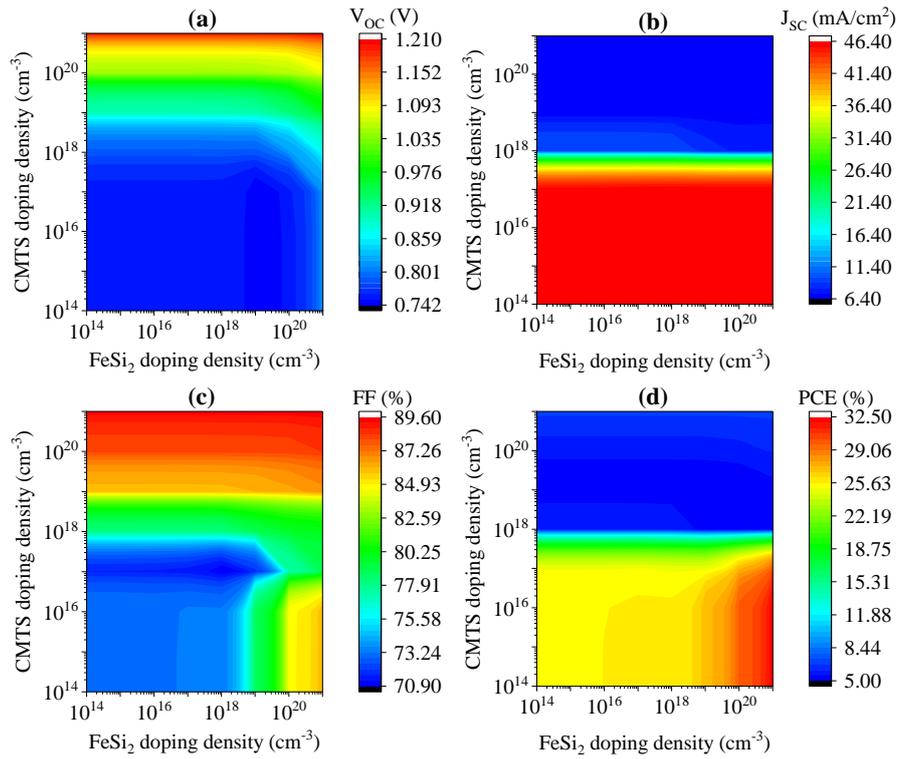

Fig. 7. Combined effect of CMTS and FeSi$_2$ doping densities on $V_{OC}$, $J_{SC}$, FF, and PCE of the device. All other material parameters were kept fixed at their default values, as listed in Tables 1 and 2.

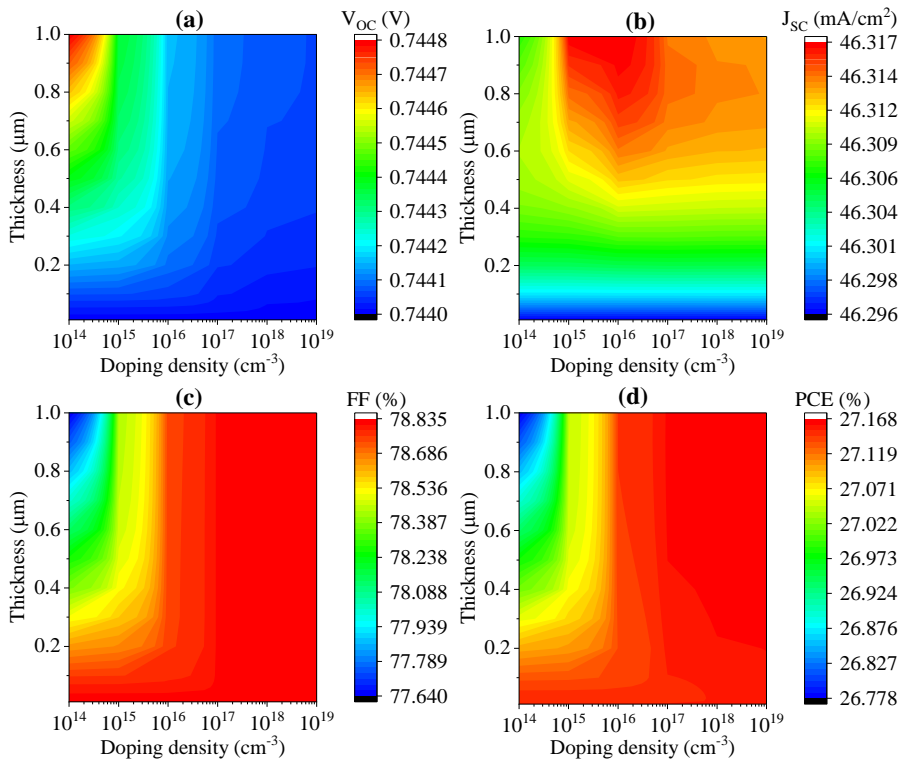

Fig. 8. Effect of thickness and doping density of the ETL (CdS) on $V_{OC}$, $J_{SC}$, FF, and PCE of the device. All other material parameters were kept fixed at their default values, as listed in Tables 1 and 2.

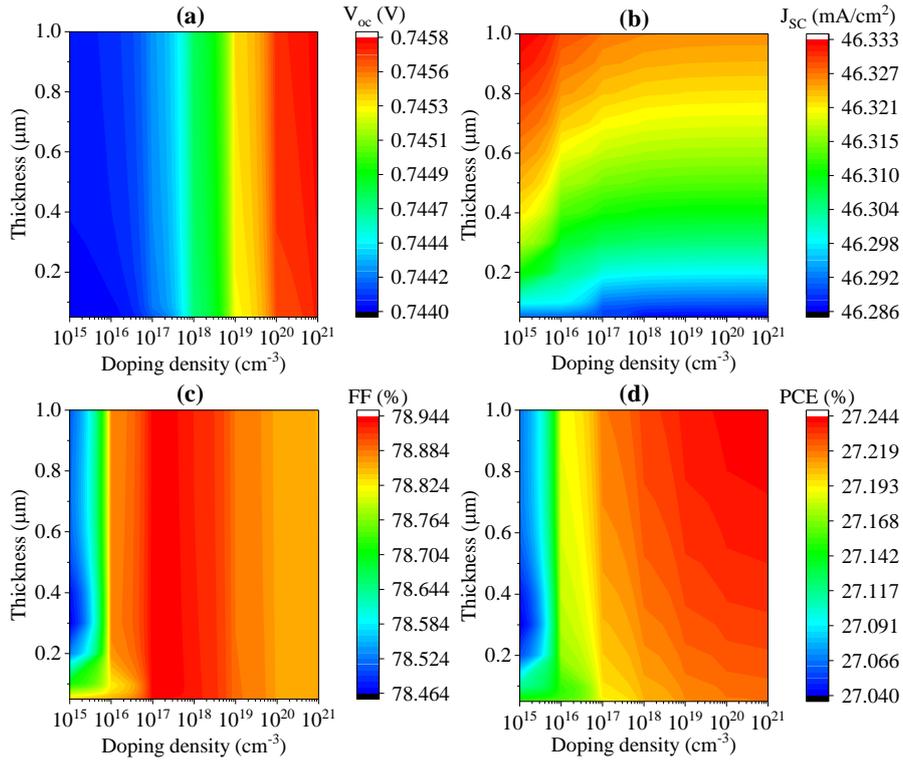

Fig. 9. Effect of thickness and doping density of the HTL ($Cu_2O$) on $V_{OC}$, $J_{SC}$, FF, and PCE of the device. All other material parameters were kept fixed at their default values, as listed in Tables 1 and 2.

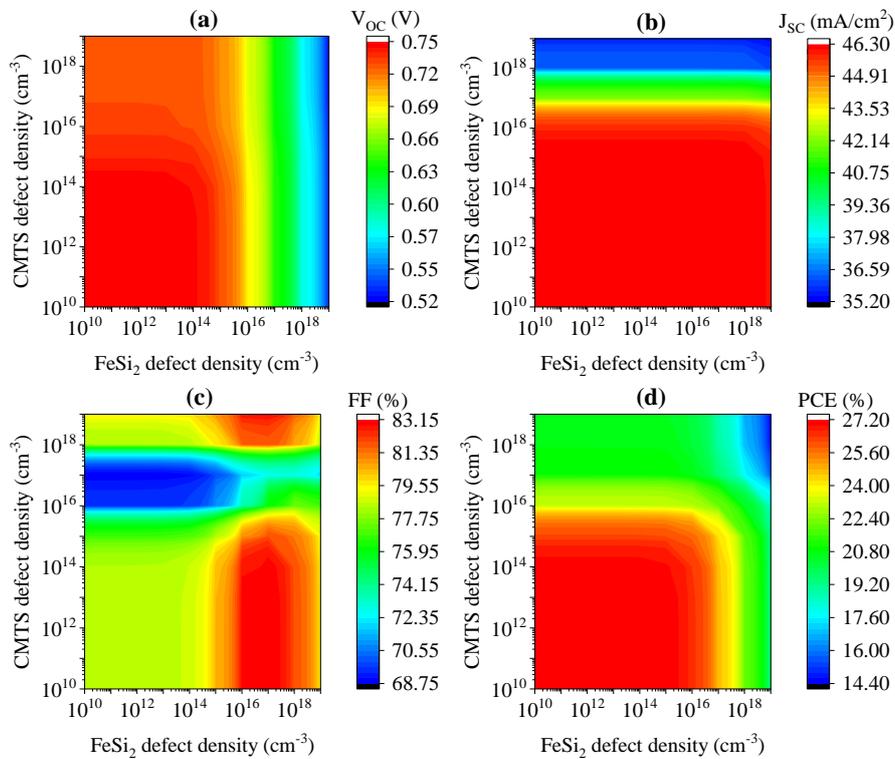

Fig. 10. Effect of bulk defect density in CMTS and $FeSi_2$ on $V_{OC}$, $J_{SC}$, FF, and PCE of the device. All other material parameters were kept fixed at their default values, as listed in Tables 1 and 2.

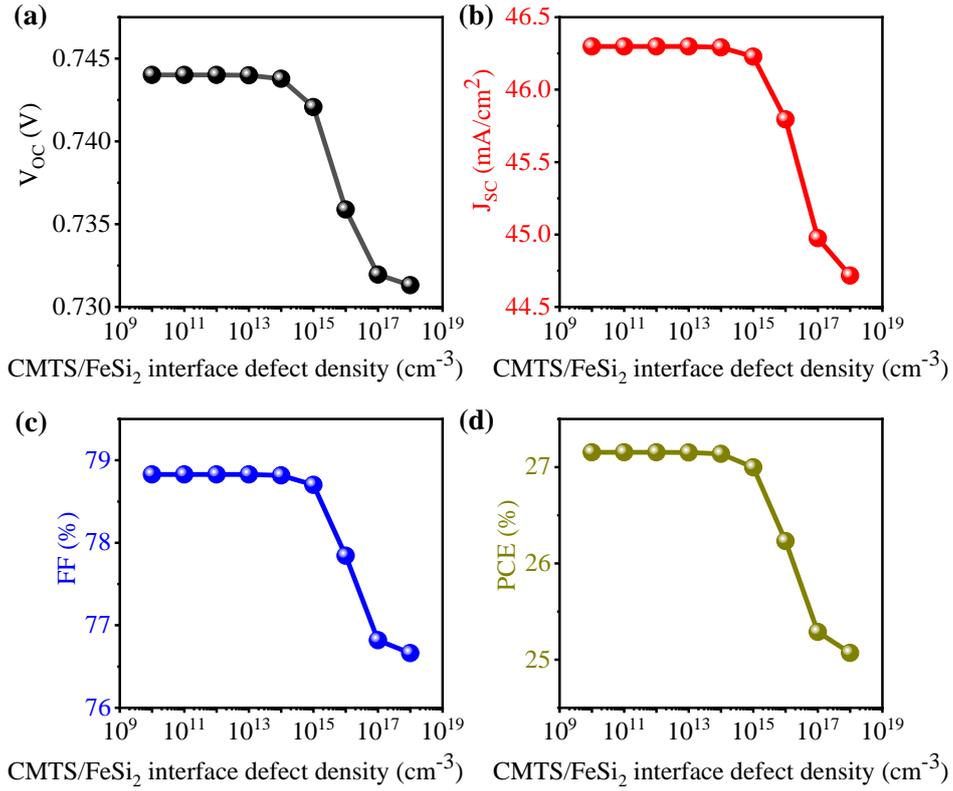

Fig. 11. Effect of defect density at CMTS/FeSi$_2$ interface on V$_{OC}$, J$_{SC}$, FF, and PCE of the device. All other material parameters were kept fixed at their default values, as listed in Tables 1 and 2.

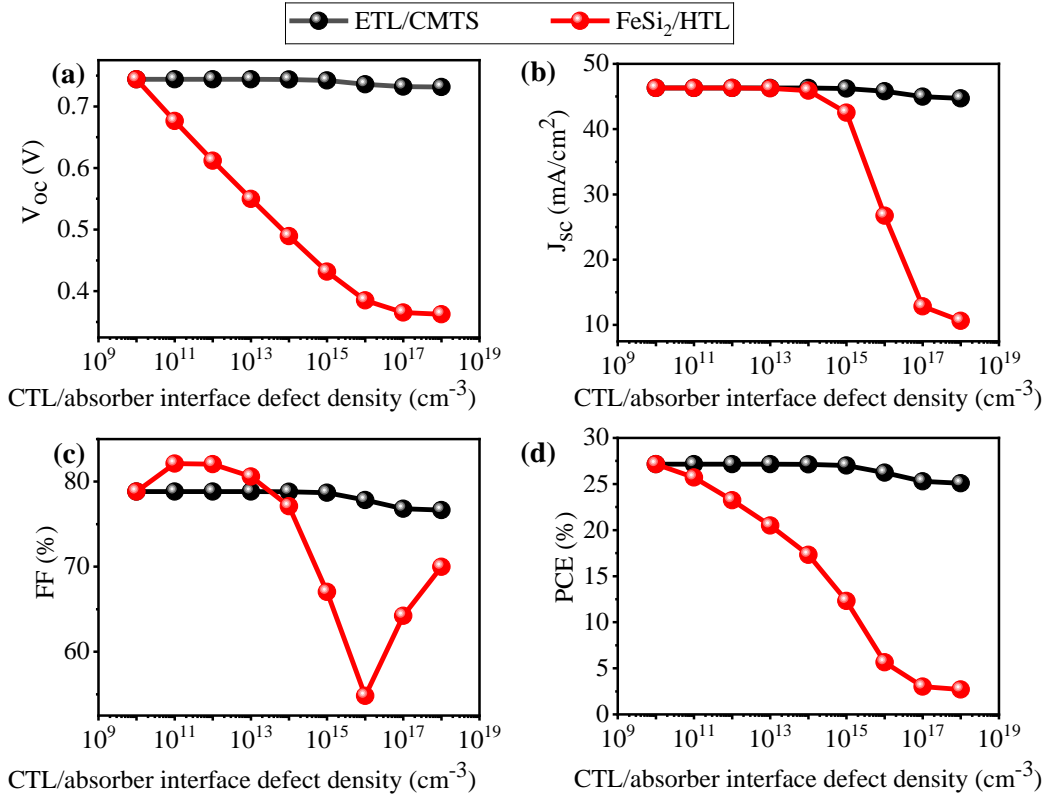

Fig. 12. Effect of defect density at charge transport layer (CTL) and absorber interfaces (ETL/CMTS and FeSi$_2$/HTL interfaces) on V$_{OC}$, J$_{SC}$, FF, and PCE of the device. All other material parameters were kept fixed at their default values, as listed in Tables 1 and 2.

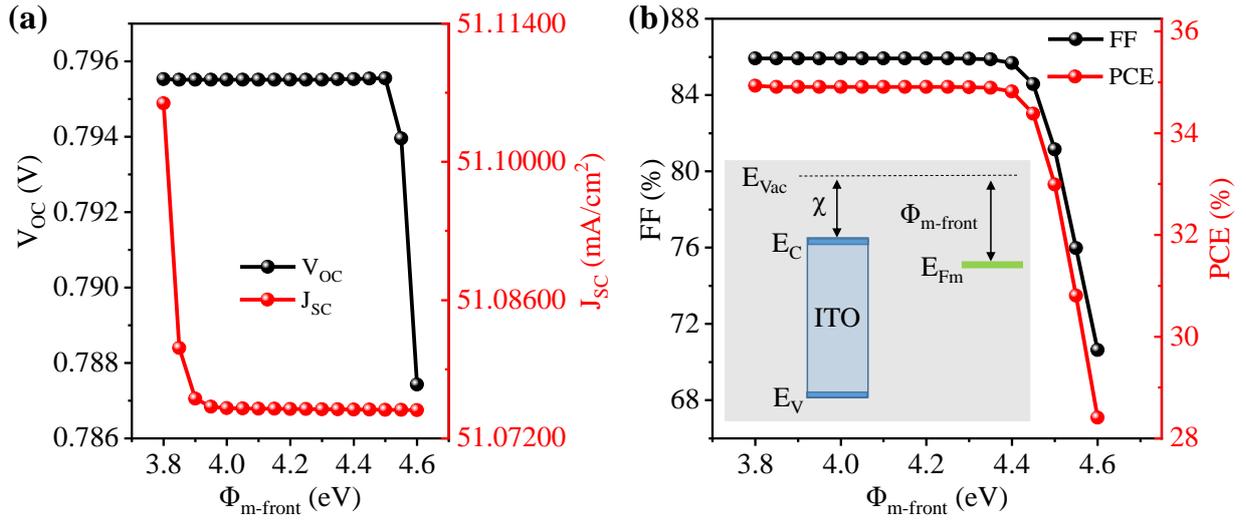

Fig. 13. Effect of front contact metal work function ($\Phi_{m\text{-front}}$) on $V_{OC}$, $J_{SC}$, FF, and PCE of the device. All other material parameters were kept fixed at their default values, as listed in Tables 1 and 2. The energy band alignment between the front contact metal and the adjacent ITO layer is schematically shown under the gray box in (b), for a clearer visualization of the explanations discussed in section 4.10. $E_{Vac}$ denotes the vacuum energy level, $E_C$ and $E_V$ represent the conduction and valence band edges of ITO, respectively, $\chi$ is the electron affinity of ITO, $E_{Fm}$ is the Fermi energy level of the front contact metal. The parameters shown inside the gray box are not drawn to scale.

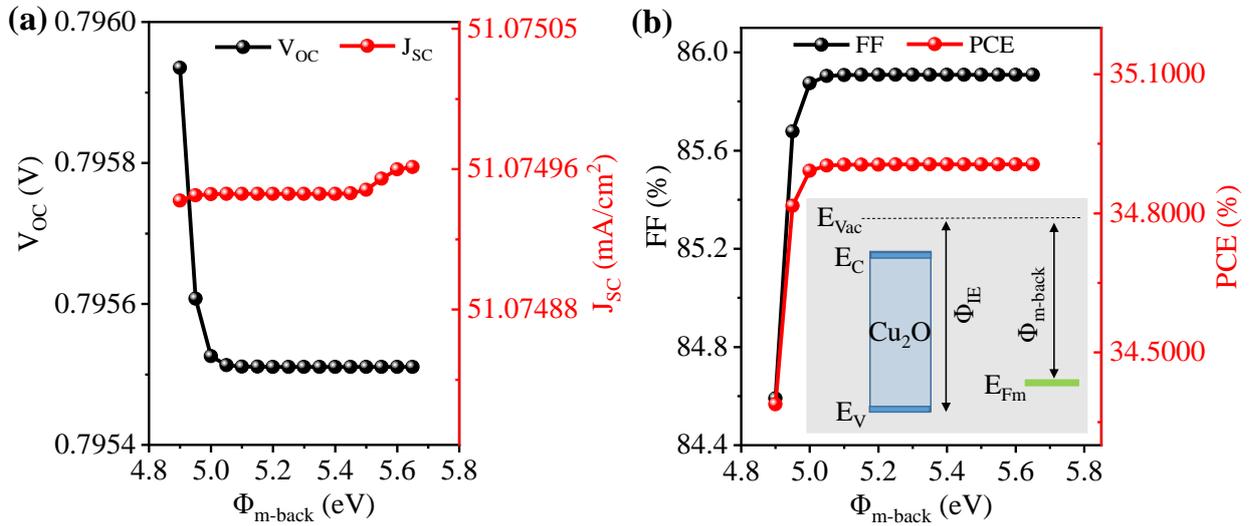

Fig. 14. Effect of back contact metal work function ($\Phi_{m\text{-back}}$) on $V_{OC}$, $J_{SC}$, FF, and PCE of the device. All other material parameters were kept fixed at their default values, as listed in Tables 1 and 2. The energy band alignment between the back contact metal and the adjacent $Cu_2O$ layer is schematically shown under the gray box in (b), for a clearer visualization of the explanations discussed in section 4.10. $E_{Vac}$ denotes the vacuum energy level, $E_C$ and $E_V$ represent the conduction and valence band edges of $Cu_2O$, respectively, $\Phi_{IE}$ is the ionization energy of $Cu_2O$, $E_{Fm}$ is the Fermi energy level of the back contact metal. The parameters shown inside the gray box are not drawn to scale.

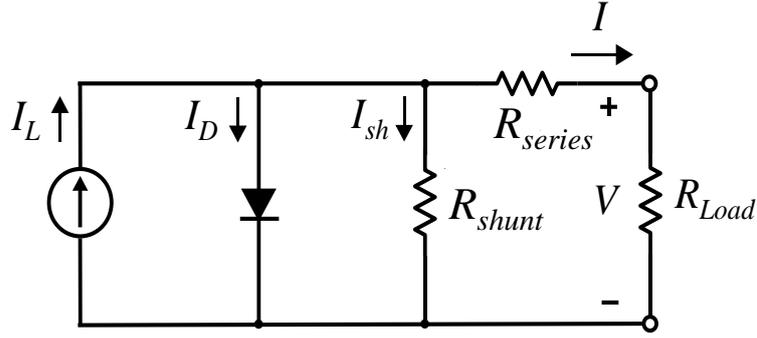

Fig. 15. Equivalent circuit representation of a solar cell. $I$ denotes the output current delivered to the external load ($R_{Load}$), $I_L$ represents the photocurrent, $I_D$ is the dark current, and $I_{sh}$ corresponds to the current flowing through the shunt branch. $V$ is the terminal voltage across the load, while $R_{series}$ and $R_{shunt}$ indicate the series and shunt resistances, respectively.

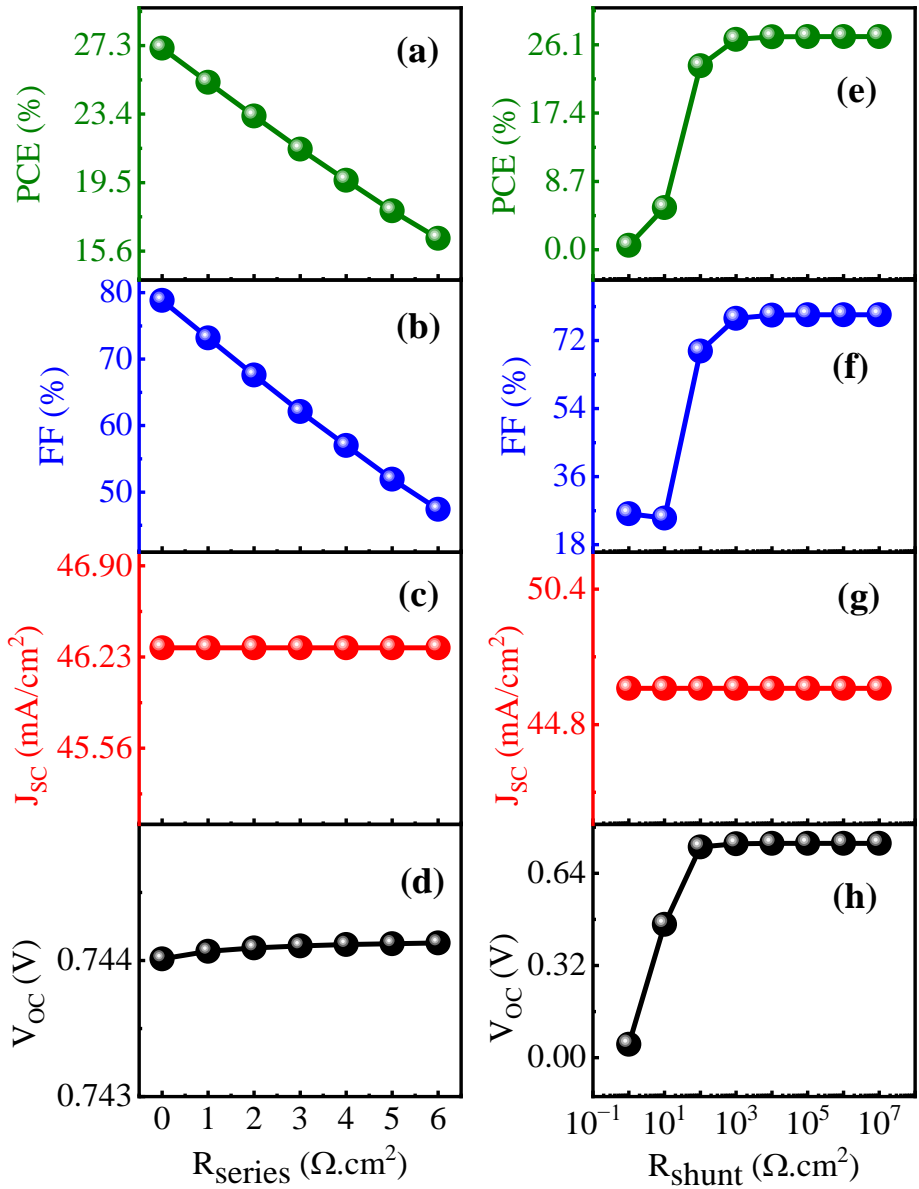

Fig. 16. Effect of series and shunt resistance ($R_{series}$ and $R_{shunt}$ respectively) on $V_{OC}$, $J_{SC}$, FF, and PCE of the device. All other material parameters were kept fixed at their default values, as listed in Tables 1 and 2.

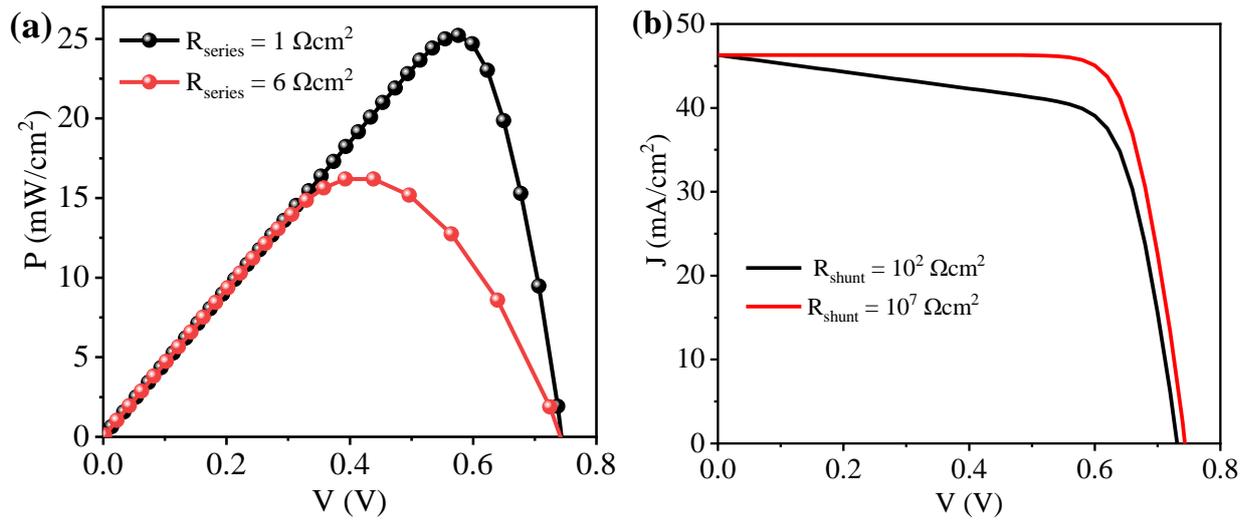

Fig. 17. (a) Power (P) versus voltage for two different series resistances and (b) J-V characteristics for two different shunt resistances.

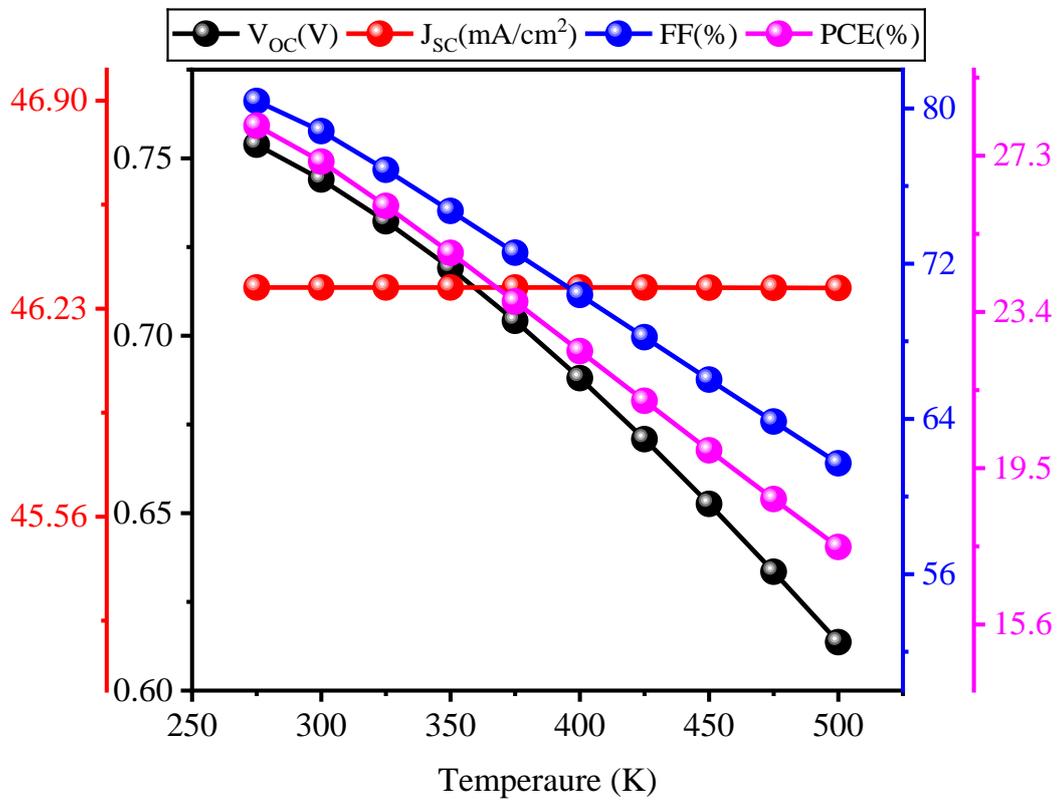

Fig. 18. Effect of temperature (T) on $V_{OC}$, $J_{SC}$, FF, and PCE of the device. All other material parameters were kept fixed at their default values, as listed in Tables 1 and 2.

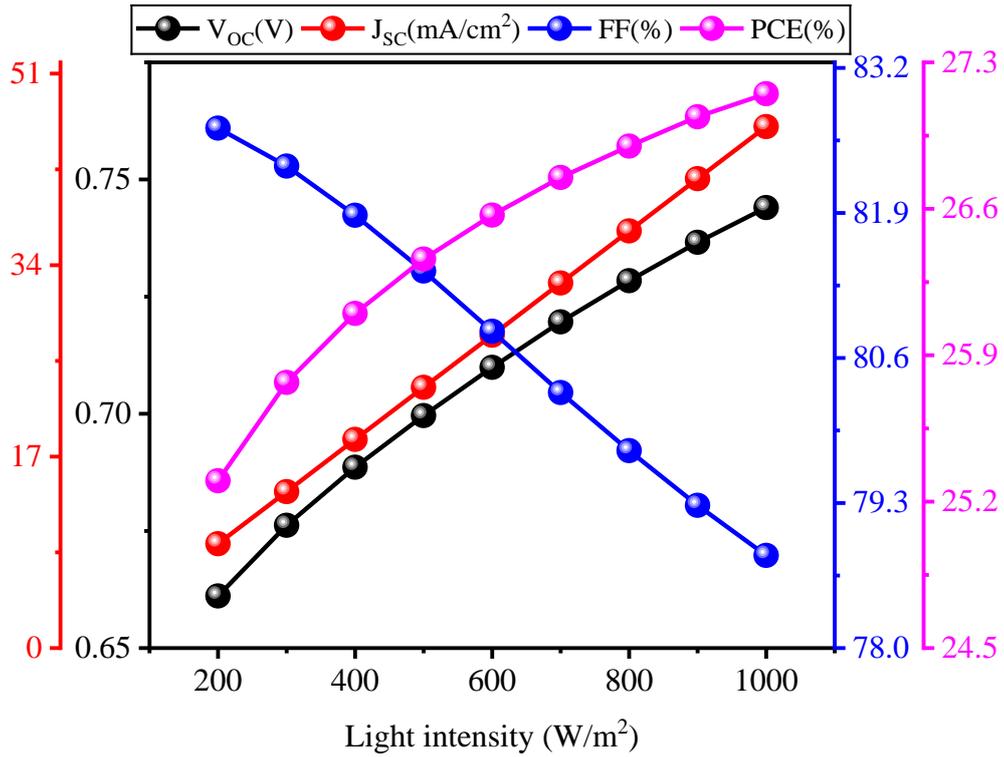

Fig. 19. Effect of incident light intesity on $V_{OC}$, $J_{SC}$, FF, and PCE of the device. All other material parameters were kept fixed at their default values, as listed in Tables 1 and 2.

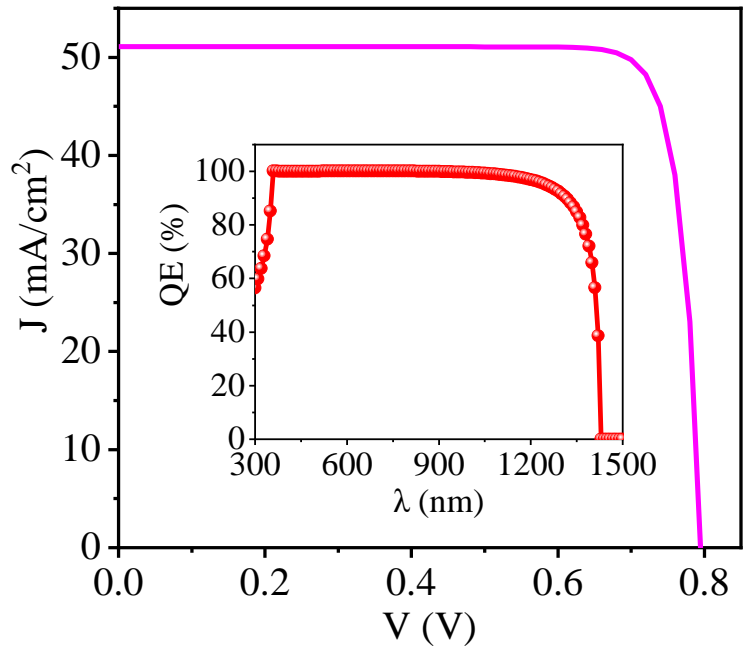

Fig. 20. J-V characteristics of the optimized dual-absorber solar cell. Inset shows the corresponding QE versus wavelength curve. Thickness of different layers were set as: $Cu_2O$ = 1 μm, $FeSi_2$ = 0.8 μm, CMTS = 0.1 μm, CdS = 1 μm, and ITO = 0.3 μm; doping densities were set as: $Cu_2O$ = $10^{21}$ cm$^{-3}$, $FeSi_2$ = $10^{21}$ cm$^{-3}$, CMTS = $10^{14}$ cm$^{-3}$, CdS = $10^{21}$ cm$^{-3}$, and ITO = $10^{21}$ cm$^{-3}$. All other material parameters were kept fixed at their default values, as listed in Tables 1 and 2.